\documentclass[aps,prd,noshowpacs,preprintnumbers,nofootinbib,floatfix,twoside]{revtex4}   

\usepackage{graphicx}  
\usepackage{amsmath}   
\usepackage{amssymb}   

\def\eq#1{{Eq.~(\ref{#1})}}

\newcommand{\cc}{cosmological constant}
\newcommand{\del}{\partial}
\newcommand{\dphi}{\partial_i \phi \partial^i \phi}

\newcommand{\Cal}[1]{\ensuremath{\mathcal{#1}}}
\newcommand{\dV}{\ensuremath{\partial\Cal{V}}}
\newcommand{\LL}{Lanczos-Lovelock }
\newcommand{\D}{\ensuremath{\nabla}}
\newcommand{\Riem}[4]{\ensuremath{R^{#1 #2}_{#3 #4}}}
\newcommand{\Alt}[6]{\ensuremath{\delta^{#1 #2 ... #3}_{#4 #5
      ... #6}}} 

\newcommand{\AltC}[8]{\ensuremath{\delta^{#1 #2 #3... #4}_{#5 #6 #7
      ... #8}}}  
\newcommand{\sD}[1]{\sum_{m=1}^{K}{#1}}
\newcommand{\LDm}{\ensuremath{\Cal{L}^{(D)}_m}}
\newcommand{\eqn}[1]{Eq.\eqref{#1}}
\newcommand{\ph}[1]{\phantom{#1}}

\def\frab#1#2{\left(\frac{#1}{#2}\right)}

\begin{document}

\title{DARK ENERGY AND GRAVITY}

\author{T. Padmanabhan}
\email{paddy@iucaa.ernet.in}
\affiliation{IUCAA, Post Bag 4, Ganeshkhind, Pune - 411 007, India\\}

\date{\today}

\begin{abstract}
  I review the problem of dark energy  focussing on \cc\ as the candidate and discuss what it tells us regarding the nature of gravity. Part 1  briefly overviews the currently popular `concordance cosmology' and  summarises the evidence for dark energy. It also provides the observational and theoretical arguments in favour of the \cc\ as a candidate and emphasises why no other approach really solves the conceptual problems usually attributed to \cc . 
 
 Part 2 describes some of the approaches to understand the nature of the \cc\ and attempts to extract certain key ingredients which must be present in any viable solution. In the conventional
approach, the equations of motion for matter fields are invariant under the shift of the matter
Lagrangian by a constant while gravity breaks this symmetry. I argue that
until the gravity is made to respect this symmetry, one cannot obtain a satisfactory solution
to the cosmological constant problem. \textit{Hence \cc\ problem essentially has to do with our understanding of the nature of gravity}. 

Part 3 discusses such an alternative  perspective on gravity in which the gravitational interaction -- described   
in terms of a metric on a smooth spacetime -- is an emergent,
long wavelength phenomenon,   and can be described in terms of an effective theory using an action associated with  normalized vectors in the spacetime. This action is explicitly invariant under the shift of the matter energy momentum tensor $T_{ab}\to T_{ab}+\Lambda g_{ab}$ and any bulk \cc\ can be gauged away. Extremizing this action leads to an equation \textit{determining the background geometry} which gives Einstein's theory at the lowest order with \LL type corrections.   In this approach, the observed value of the \cc\ has to arise from the energy fluctuations of degrees of freedom located in the boundary of a spacetime region. 

 \end{abstract}
 
\maketitle
\tableofcontents
\section{Cosmological Constant as the dark energy}
 
\subsection{The cosmological paradigm}

A host of different observations, which became available in the last couple of decades, have thrusted upon us a  preposterous
composition for the energy density of different components in the universe which defies any simple explanation.
The energy densities of the different species which drive the expansion of the universe, can be measured in terms of a \textit{critical energy density} $\rho_c=3H^2_0/8\pi G$  where $H_0=(\dot a/a)_0$
is the rate of expansion of the universe at present.  The variables $\Omega_i=\rho_i/\rho_c$ 
will then give the fractional contribution of different components of the universe ($i$ denoting baryons, dark matter, radiation, etc.) to the  critical density required to close the universe. Observations  suggest that the universe has
 $0.98\lesssim\Omega_{tot}\lesssim1.08$  with radiation (R), baryons (B), dark matter, made of weakly interacting massive particles (DM) and dark energy (DE) contributing  $\Omega_R\simeq 5\times 10^{-5},\Omega_B\simeq 0.04,\Omega_{DM}\simeq 0.26,\Omega_{DE}\simeq 0.7,$ respectively. All known observations \cite{cmbr,baryon,h}
are consistent with such an --- admittedly weird --- composition for the universe.

The conventional cosmological paradigm --- which is remarkably successful --- is based on these numbers and can be summarised \cite{adcos} as follows:
The  key idea is that if there existed small fluctuations in the energy density in the early universe, then gravitational instability can amplify them in a well-understood manner  leading to structures like galaxies etc. today. The most popular model for generating these fluctuations is based on the idea that if the very early universe went through an inflationary phase \cite{inflation}, then the quantum fluctuations of the field driving the inflation can lead to energy density fluctuations \cite{genofpert,tplp}. It is possible to construct models of inflation such that these fluctuations are described by a Gaussian random field and are characterized by a power spectrum of the form $P(k)=A k^n$ with $n\simeq 1$. The inflationary models cannot predict the value of the amplitude $A$ in an unambiguous manner. But it can be determined from CMBR observations and the inflationary model parameters can be fine-tuned to reproduce the observed value. The CMBR observations are consistent with the inflationary model for the generation of perturbations and gives $A\simeq (28.3 h^{-1} Mpc)^4$ and $n\lesssim 1$. (The first results were from COBE \cite{cobeanaly} and
WMAP has re-confirmed them with far greater accuracy).
When the perturbation is small, one can use well defined linear perturbation theory to study its growth. But when $\delta\approx(\delta\rho/\rho)$ is comparable to unity the perturbation theory
breaks down. Since there is more power at small scales, smaller scales go non-linear first and structure forms hierarchically. 
The non linear evolution of the  \textit{dark matter halos}  can be understood by simulations 
 as well as theoretical models based on approximate 
 ansatz \cite{nlapprox} and  nonlinear scaling relations \cite{nsr}.
 The baryons in the halo will cool and undergo collapse
 in a fairly complex manner because of gas dynamical processes. 
 It seems unlikely that the baryonic collapse and galaxy formation can be understood
 by analytic approximations; one needs to do high resolution computer simulations
 to make any progress \cite{baryonsimulations}. The results obtained from all these
 attempts are broadly consistent with observations but the summary given above demonstrates that modeling the universe and comparing the theory with observations is a rather involved affair.
 
 So, to the zeroth order, the universe is characterized by just seven numbers: $h\approx 0.7$ describing the current rate of expansion; $\Omega_{DE}\simeq 0.7,\Omega_{DM}\simeq 0.26,\Omega_B\simeq 0.04,\Omega_R\simeq 5\times 10^{-5}$ giving the composition of the universe; the amplitude $A\simeq (28.3 h^{-1} Mpc)^4$ and the index $n\simeq 1$ of the initial perturbations. 
 The remaining challenge, of course,  is to make some sense out of these numbers themselves from a more fundamental point of view. Among all these components, the dark energy, which exerts negative pressure, is probably the weirdest and --- since non-cosmologists often wonder how strong is the evidence for it --- it is useful keep the following points in mind:
 \begin{itemize}
\item 
The rapid acceptance of dark energy by the community is partially due to the fact that --- even before the supernova data came up --- there were strong indications for the existence of dark energy.  Early analysis of several observations \cite{earlyde}
indicated that this component is unclustered and has negative pressure.  This is, of course, confirmed dramatically by the supernova observations\cite{sn,snls}. (For a critical look at the current data, see \cite{tptirthsn1}; a sample of recent SN data analysis papers can be found in ref. \cite{sndataanalysis}.)  
\item
The WMAP-CMBR data with a reasonable prior on Hubble constant implies $\Omega_{tot}\approx 1$ while a host of other astronomical observations show that the clustered matter contributes only about
$\Omega_{DM}\approx 0.25-0.4$. Together, they require a unclustered (negative pressure) component in the universe independent of SN data. \textit{It, therefore, seems very unlikely that dark energy will ``go away".}

\end{itemize}

The key observational feature of dark energy is that --- treated as a fluid with a stress tensor $T^a_b=$ dia     $(\rho, -p, -p,-p)$ 
--- it has an equation state $p=w\rho$ with $w \lesssim -0.8$ at the present epoch. 
The spatial part  ${\bf g}$  of the geodesic acceleration (which measures the 
  relative acceleration of two geodesics in the spacetime) satisfies an \textit{exact} equation
  in general relativity  given by:
  \begin{equation}
  \nabla \cdot {\bf g} = - 4\pi G (\rho + 3p)
  \label{nextnine}
  \end{equation} 
 This  shows that the source of geodesic  acceleration is $(\rho + 3p)$ and not $\rho$.
  As long as $(\rho + 3p) > 0$, gravity remains attractive while $(\rho + 3p) <0$ can
  lead to `repulsive' gravitational effects. In other words, dark energy with sufficiently negative pressure will
  accelerate the expansion of the universe, once it starts dominating over the normal matter.  This is precisely what is established from the study of high redshift supernova, which can be used to determine the expansion
rate of the universe in the past \cite{sn,snls}.

The simplest model for  a fluid with negative pressure is the
cosmological constant (for a sample of recent reviews, see ref. \cite{cc}) with $w=-1,\rho =-p=$ constant.
If  dark energy is indeed the cosmological constant, then it introduces a fundamental length scale in the theory $L_\Lambda\equiv H_\Lambda^{-1}$, related to the constant dark energy density $\rho_{_{\rm DE}}$ by 
$H_\Lambda^2\equiv (8\pi G\rho_{_{\rm DE}}/3)$.
In classical general relativity,
    based on  $G, c $ and $L_\Lambda$,  it
  is not possible to construct any dimensionless combination from these constants. But when one introduces the Planck constant, $\hbar$, it is  possible
  to form the dimensionless combination $H^2_\Lambda(G\hbar/c^3) \equiv  (L_P^2/L_\Lambda^2)$.
  Observations then require $(L_P^2/L_\Lambda^2) \lesssim 10^{-123}$.
  This will require enormous fine tuning. What is more,
 in the past, the energy density of 
  normal matter and radiation  would have been higher while the energy density contributed by the  cosmological constant
  does not change.  Hence we need to adjust the energy densities
  of normal matter and cosmological constant in the early epoch very carefully so that
  $\rho_\Lambda\gtrsim \rho_{\rm NR}$ around the current epoch.
  This raises the second of the two cosmological constant problems:
  Why is $(\rho_\Lambda/ \rho_{\rm NR}) = \mathcal{O} (1)$ at the 
  {\it current} phase of the universe ? These are the two conventional conceptual difficulties associated with the \cc\ and have been discussed extensively in literature.

\subsection{The `denial' approach to the cosmological constant}    
  
  Because of these conceptual problems associated with the cosmological constant, people have explored a large variety of alternative possibilities. The most popular among them uses a scalar field $\phi$ with a suitably chosen potential $V(\phi)$ so as to make the vacuum energy vary with time. The hope then is that, one can find a model in which the current value can be explained naturally without any fine tuning.
  A simple form of the source with variable $w$ are   scalar fields with
  Lagrangians of different forms, of which we will discuss two possibilities:
    \begin{equation}
  \mathcal{L}_{\rm quin} = \frac{1}{2} \partial_a \phi \partial^a \phi - V(\phi); \quad \mathcal{L}_{\rm tach}
  = -V(\phi) [1-\partial_a\phi\partial^a\phi]^{1/2}
  \label{lquineq}
  \end{equation}
  Both these Lagrangians involve one arbitrary function $V(\phi)$. The first one,
  $\mathcal{L}_{\rm quin}$,  which is a natural generalization of the Lagrangian for
  a non-relativistic particle, $L=(1/2)\dot q^2 -V(q)$, is usually called quintessence (for
  a small sample of models, see \cite{phiindustry}).
    When it acts as a source in Friedman universe,
   it is characterized by a time dependent
  $w(t)$ with
    \begin{equation}
  \rho_q(t) = \frac{1}{2} \dot\phi^2 + V; \quad p_q(t) = \frac{1}{2} \dot\phi^2 - V; \quad w_q
  = \frac{1-(2V/\dot\phi^2)}{1+ (2V/\dot\phi^2)}
  \label{quintdetail}
  \end{equation}

The structure of the second Lagrangian in Eq.~(\ref{lquineq}) (which arises in string theory)  can be understood by a simple analogy from
special relativity. A relativistic particle with  (one dimensional) position
$q(t)$ and mass $m$ is described by the Lagrangian $L = -m \sqrt{1-\dot q^2}$.
It has the energy $E = m/  \sqrt{1-\dot q^2}$ and momentum $k = m \dot
q/\sqrt{1-\dot q^2} $ which are related by $E^2 = k^2 + m^2$.  As is well
known, this allows the possibility of having \textit{massless} particles with finite
energy for which $E^2=k^2$. This is achieved by taking the limit of $m \to 0$
and $\dot q \to 1$, while keeping the ratio in $E = m/  \sqrt{1-\dot q^2}$
finite.  The momentum acquires a life of its own,  unconnected with the
velocity  $\dot q$, and the energy is expressed in terms of the  momentum
(rather than in terms of $\dot q$)  in the Hamiltonian formulation. We can now
construct a field theory by upgrading $q(t)$ to a field $\phi$. Relativistic
invariance now  requires $\phi $ to depend on both space and time [$\phi =
\phi(t, {\bf x})$] and $\dot q^2$ to be replaced by $\partial_i \phi \partial^i
\phi$. It is also possible now to treat the mass parameter $m$ as a function of
$\phi$, say, $V(\phi)$ thereby obtaining a field theoretic Lagrangian $L =-
V(\phi) \sqrt{1 - \del^i \phi \del_i \phi}$. The Hamiltonian  structure of this
theory is algebraically very similar to the special  relativistic example  we
started with. In particular, the theory allows solutions in which $V\to 0$,
$\dphi \to 1$ simultaneously, keeping the energy (density) finite.  Such
solutions will have finite momentum density (analogous to a massless particle
with finite  momentum $k$) and energy density. Since the solutions can now
depend on both space and time (unlike the special relativistic example in which
$q$ depended only on time), the momentum density can be an arbitrary function
of the spatial coordinate. The structure of this Lagrangian is similar to those analyzed in a wide class of models
   called {\it K-essence} \cite{kessence} and  provides a rich gamut of possibilities in the
context of cosmology
 \cite{tptachyon,tachyon}.

   Since  the quintessence field (or the tachyonic field)   has
   an undetermined free function $V(\phi)$, it is possible to choose this function
  in order to produce a given expansion history of the universe characterized by the function $H(a)=\dot a/a$ expressed in terms of $a$. 
  To see this explicitly, let
   us assume that the universe has two forms of energy density with $\rho(a) =\rho_{\rm known}
  (a) + \rho_\phi(a)$ where $\rho_{\rm known}(a)$ arises from any known forms of source 
  (matter, radiation, ...) and
  $\rho_\phi(a) $ is due to a scalar field.  
  Let us first consider quintessence. Here,  the potential is given implicitly by the form
  \cite{ellis,tptachyon}
  \begin{equation}
  V(a) = \frac{1}{16\pi G} H (1-Q)\left[6H + 2aH' - \frac{aH Q'}{1-Q}\right]
  \label{voft}
   \end{equation} 
    \begin{equation}
    \phi (a) =  \left[ \frac{1}{8\pi G}\right]^{1/2} \int \frac{da}{a}
     \left[ aQ' - (1-Q)\frac{d \ln H^2}{d\ln a}\right]^{1/2}
    \label{phioft}
    \end{equation} 
   where $Q (a) \equiv [8\pi G \rho_{\rm known}(a) / 3H^2(a)]$ and prime denotes differentiation with respect to $a$.
   Given any
   $H(a),Q(a)$, these equations determine $V(a)$ and $\phi(a)$ and thus the potential $V(\phi)$. 
   \textit{Every quintessence model studied in the literature can be obtained from these equations.}
  
  Similar results exists for the tachyonic scalar field as well \cite{tptachyon}. For example, given
  any $H(a)$, one can construct a tachyonic potential $V(\phi)$ so that the scalar field is the 
  source for the cosmology. The equations determining $V(\phi)$  are now given by:
  \begin{equation}
  \phi(a) = \int \frac{da}{aH} \left(\frac{aQ'}{3(1-Q)}
   -{2\over 3}{a H'\over H}\right)^{1/2}
  \label{finalone}
  \end{equation}
   \begin{equation}
   V(a) = {3H^2 \over 8\pi G}(1-Q) \left( 1 + {2\over 3}{a H'\over H}-\frac{aQ'}{3(1-Q)}\right)^{1/2}
   \label{finaltwo}
   \end{equation}
   Equations (\ref{finalone}) and (\ref{finaltwo}) completely solve the problem. Given any
   $H(a)$, these equations determine $V(a)$ and $\phi(a)$ and thus the potential $V(\phi)$. 
A wide variety of phenomenological models with time dependent
  \cc\ have been considered in the literature; all of these can be 
   mapped to a 
  scalar field model with a suitable $V(\phi)$.

  While the scalar field models enjoy considerable popularity (one reason being they are easy to construct!)
  it is very doubtful whether they have helped us to understand the nature of the dark energy
  at any deeper level. These
  models, viewed objectively, suffer from several shortcomings:
   \begin{figure}[ht]
 \includegraphics[scale=0.5]{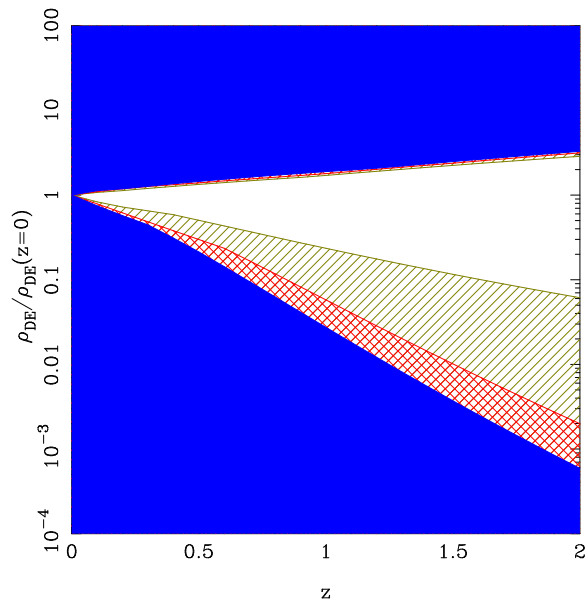}
 \caption{
 The observational constraints on the variation of dark energy density  as a function of
 redshift  from WMAP and SNLS data (see \cite{jbp}). The green/hatched region is
 excluded  at $68\%$ confidence limit, red/cross-hatched region at $95\%$
 confidence level  and the blue/solid region at $99\%$ confidence limit. The
 white region shows the allowed range of variation of dark energy at $68\%$
 confidence limit.  
  }
 \label{fig:bjp2ps}
 \end{figure} 
 \begin{itemize}
  \item
  They have no predictive power. As explicitly demonstrated above, virtually every form of $a(t)$ can be modeled by a suitable ``designer" $V(\phi)$.
  \item
  These models are  degenerate in another sense. The previous discussion  illustrates that even when $w(a)$ is known/specified, it is not possible to proceed further and determine
  the nature of the scalar field Lagrangian. The explicit examples given above show that there
  are {\em at least} two different forms of scalar field Lagrangians --- corresponding to
  the quintessence or the tachyonic field --- which could lead to
  the same $w(a)$. (See the first paper in ref.\cite{tptirthsn1} for an explicit example of such a construction.)
  
  \item
  By and large, the potentials  used in the literature have no natural field theoretical justification. All of them are non-renormalisable in the conventional sense and have to be interpreted as a low energy effective potential in an ad hoc manner.
  \item
  One key difference between \cc\ and scalar field models is that the latter lead to a $w(a)$ which varies with time. If observations have demanded this, or even if observations have ruled out $w=-1$ at the present epoch,
  then one would have been forced to take alternative models seriously. However, all available observations are consistent with \cc\ ($w=-1$) and --- in fact --- the possible variation of $w$ is strongly constrained \cite{jbp} as shown in Figure \ref{fig:bjp2ps}. 
  \item
  While on the topic of observational constraints on $w(t)$, the following point needs to be stressed: One should be careful about the hidden assumptions in the statistical analysis of these data. Claims regarding the value of $w$ depends crucially on the data sets used, priors which are assumed and possible parameterizations which are adopted. (For more details related to these issues, see the last reference in \cite{jbp}.) It is fair to say that all currently available data is consistent with $w=-1$. Further, there is some amount of tension between WMAP and SN-Gold data with the recent SNLS data \cite{snls} being more concordant with WMAP than the SN Gold data. 
  \item
  The most serious problem with the scalar field models is the following: All the scalar field potentials require fine tuning of the parameters in order to be viable. This is obvious in the quintessence models in which adding a constant to the potential is the same as invoking a \cc. So to make the quintessence models work, \textit{we first need to assume the \cc\ is zero.} These models, therefore, merely push the cosmological constant problem to another level, making it somebody else's problem!.
  
 \end{itemize}
 
The last point makes clear that if we shift $\mathcal{L}\to \mathcal{L}_{\rm matt} - 2\lambda_m$ in an otherwise successful scalar field model for dark energy, we end up `switching on' the cosmological constant and raising the problems again. It is therefore important to address this issue, which we will discuss in Part 3.
 
 Given this situation, we shall first take a more serious look at the \cc\ as the source of dark energy in the universe.  

\section{Aspects of the Cosmological Constant}

\subsection{Facing up to the Challenge}
 
The observational and theoretical features described above suggests that one should consider \cc\ as the most natural candidate for dark energy. Though it leads to well known problems, it is also the most economical [just one number] and simplest  explanation for all the observations. 

Once we invoke the \cc, classical gravity will be described by the three constants $G,c$ and $\Lambda\equiv L_\Lambda^{-2}$. Since $\Lambda(G\hbar/c^3)\equiv (L_P/L_\Lambda)^2\approx 10^{-123}$, it is obvious that the \cc\ is telling us something regarding \textit{quantum gravity}, indicated by the combination $G\hbar$. \textit{An acid test for any quantum gravity model will be its ability to explain this value;} needless to say, all the currently available models --- strings, loops etc.  --- flunk this test. Even assuming that this is more of an  issue in semiclassical gravity rather than quantum gravity, one cannot help noticing that several different approaches to semiclassical gravity \cite{semicgrav} are silent about cosmological constant.

In terms of the energy scales, the \cc\ problem is an infra red problem \textit{par excellence}.
At the same time, the occurrence of $\hbar$ in $\Lambda(G\hbar/c^3)$ shows that it is a relic of a quantum gravitational effect (or principle) of unknown nature. One is envisaging here a somewhat unusual possibility of  a high energy phenomenon leaving a low energy relic and an analogy will be helpful to illustrate this idea \cite{choices}. Suppose we solve the Schrodinger equation for the Helium atom for the quantum states of the two electrons $\psi(x_1,x_2)$. When the result is compared with observations, we will find that only half the states --- those in which  $\psi(x_1,x_2)$ is antisymmetric under $x_1\longleftrightarrow x_2$ interchange --- are realized in nature. But the low energy Hamiltonian for electrons in the Helium atom has no information about
this effect! Here is a low energy (IR) effect which is a relic of relativistic quantum field theory (spin-statistics theorem) that is  totally non perturbative, in the sense that writing corrections to the Hamiltonian of the Helium atom  in some $(1/c)$ expansion will {\it not} reproduce this result. I suspect the current value of \cc\ is related to quantum gravity in a similar spirit. There must exist a deep principle in quantum gravity which leaves its non-perturbative trace even in the low energy limit
that appears as the \cc.

\subsubsection{Cosmology with two length scales}

  Given the two length scales $L_P$ and $L_\Lambda$, one can construct two energy scales
 $\rho_{_{\rm UV}}=1/L_P^4$ and $\rho_{_{\rm IR}}=1/L_\Lambda^4$ in natural units ($c=\hbar=1$). There is sufficient amount of justification from different theoretical perspectives
 to treat $L_P$ as the zero point length of spacetime \cite{zeropoint}, giving a natural interpretation to $\rho_{_{\rm UV}}$. The second one, $\rho_{_{\rm IR}}$ also has a natural interpretation. Since the universe  dominated by a \cc\ at late times will be  asymptotically DeSitter with $a(t)\propto \exp (t/L_\Lambda) $ at late times, it will have a horizon and associated thermodynamics \cite{ghds} with a  temperature
 $T=H_\Lambda/2\pi$. The corresponding thermal energy density is $\rho_{thermal}\propto T^4\propto 1/L_\Lambda^4=
 \rho_{_{\rm IR}}$. Thus $L_P$ determines the \textit{highest} possible energy density in the universe while $L_\Lambda$
 determines the {\it lowest} possible energy density in this universe. As the energy density of normal matter drops below this value, $\rho_{IR}$, the thermal ambience of the DeSitter phase will remain constant and provide the irreducible `vacuum noise'. The observed dark energy density is the the geometric mean 
\begin{equation}
 \rho_{_{\rm DE}}=\sqrt{\rho_{_{\rm IR}}\rho_{_{\rm UV}}}=\frac{1}{L_P^2L_\Lambda^2}
 \label{geomean}
\end{equation} 
 of these two energy densities. If we define a dark energy length scale $L_{DE}$  such that $\rho_{_{\rm DE}}=1/L_{DE}^4$ then $L_{DE}=\sqrt{L_PL_\Lambda}$ is the geometric mean of the two length scales in the universe \cite{note1}. 
 
 \begin{figure}[!]
  \begin{center}
  \includegraphics[angle=-90,scale=0.5]{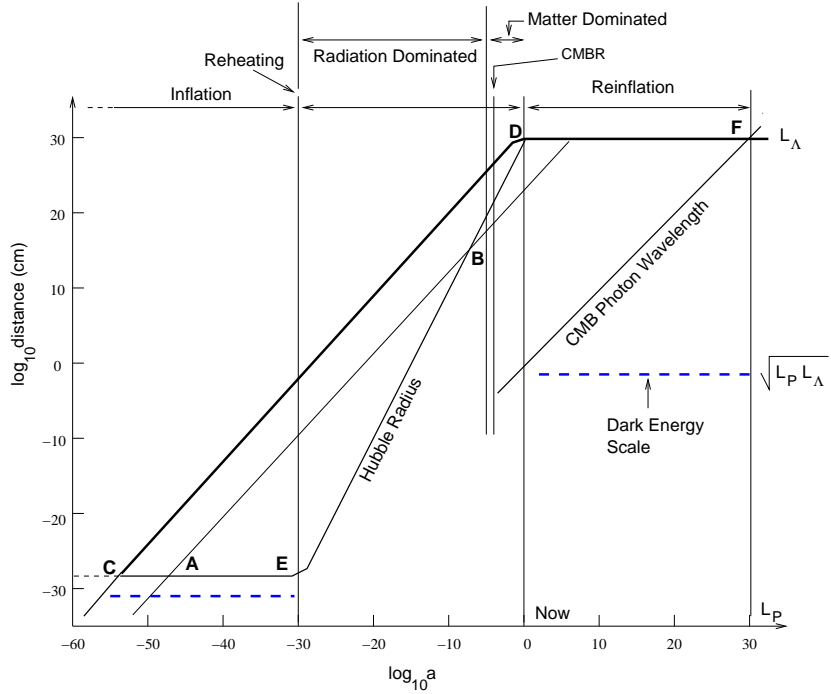}
  \end{center}
\caption{The geometrical structure of a universe with two length scales $L_P$ and $L_\Lambda$ corresponding to the Planck length and the cosmological constant \cite{plumian,bjorken}. Such a universe spends most of its time in two DeSitter phases which are (approximately) time translation invariant. The first DeSitter phase corresponds to the inflation and the second corresponds to the accelerated expansion arising from the cosmological constant. Most of the perturbations generated during the inflation will leave the Hubble radius (at some A, say) and re-enter (at B). However, perturbations which exit the Hubble radius
earlier than C will never re-enter the Hubble radius, thereby introducing a specific  dynamic range CE during the inflationary phase. The epoch F is characterized by the redshifted CMB temperature becoming equal to the DeSitter temperature $(H_\Lambda / 2\pi)$ which introduces another dynamic range DF in the accelerated expansion after which the universe is dominated by vacuum noise
of the DeSitter spacetime.}
\label{fig:tpplumian}
  \end{figure}
 
 Figure \ref{fig:tpplumian} describes some peculiar features in such a universe \cite{plumian,bjorken}. Using the characteristic length scale of expansion,
 the Hubble radius $d_H\equiv (\dot a/a)^{-1}$, we can distinguish between three different phases of such a universe. The first phase is when the universe went through a inflationary expansion with $d_H=$ constant; the second phase is the radiation/matter dominated phase in which most of the standard cosmology operates and $d_H$ increases monotonically; the third phase is that of re-inflation (or accelerated expansion) governed by the cosmological constant in which $d_H$ is again a constant. The first and last phases are time translation invariant;
 that is, $t\to t+$ constant is an (approximate) invariance for the universe in these two phases. The universe satisfies the perfect cosmological principle and is in steady state during these phases!
 
 In the most natural scenario, the two DeSitter phases (first and last) can be  of arbitrarily long duration \cite{plumian}. If  $\Omega_\Lambda\approx 0.7, \Omega_{DM}\approx 0.3$ the final DeSitter phase \textit{does} last forever; as regards the inflationary phase, nothing prevents it from lasting for arbitrarily long duration. Viewed from this perspective, the in between phase --- in which most of the `interesting' cosmological phenomena occur ---  is  of negligible measure in the span of time. It merely connects two steady state phases of the universe.
 The figure \ref{fig:tpplumian} also shows the variation of $L_{DE}$ by broken horizontal lines. 
 
 While the two DeSitter phases can last forever in principle, there is a natural cut off length scale in both of them
 which makes the region of physical relevance to be finite \cite{plumian}. Let us first discuss the case of re-inflation in the late universe. 
 As the universe grows exponentially in the phase 3, the wavelength of CMBR photons are being redshifted rapidly. When the temperature of the CMBR radiation drops below the DeSitter temperature (which happens when the wavelength of the typical CMBR photon is stretched to the $L_\Lambda$.)
 the universe will be essentially dominated by the vacuum thermal noise of the DeSitter phase.
 This happens at the point marked F when the expansion factor is $a=a_F$ determined by the
  equation $T_0 (a_0/a_{F}) = (1/2\pi L_\Lambda)$. Let $a=a_\Lambda$ be the epoch at which
  cosmological constant started dominating over matter, so that $(a_\Lambda/a_0)^3=
  (\Omega_{DM}/\Omega_\Lambda)$. Then we find that the dynamic range of 
 DF is 
 \begin{equation}
\frac{a_F}{a_\Lambda} = 2\pi T_0 L_\Lambda \left( \frac{\Omega_\Lambda}{\Omega_{DM}}\right)^{1/3}
\approx 3\times 10^{30}
\end{equation}

 One can also impose a similar bound on the physically relevant duration of inflation. 
 We know that the quantum fluctuations generated during this inflationary phase could act as seeds of structure formation in the universe \cite{genofpert}. Consider a perturbation at some given wavelength scale which is stretched with the expansion of the universe as $\lambda\propto a(t)$.
 (See the line marked AB in Figure \ref{fig:tpplumian}.)
 During the inflationary phase, the Hubble radius remains constant while the wavelength increases, so that the perturbation will `exit' the Hubble radius at some time (the point A in Figure \ref{fig:tpplumian}). In the radiation dominated phase, the Hubble radius $d_H\propto t\propto a^2$ grows faster than the wavelength $ \lambda\propto a(t)$. Hence, normally, the perturbation will `re-enter' the Hubble radius at some time (the point B in Figure \ref{fig:tpplumian}).
 If there was no re-inflation, this will make {\it all} wavelengths re-enter the Hubble radius sooner or later.
 But if the universe undergoes re-inflation, then the Hubble radius `flattens out' at late times and some of the perturbations will {\it never} reenter the Hubble radius. The limiting perturbation which just `grazes' the Hubble radius as the universe enters the re-inflationary phase is shown by the line marked CD in Figure \ref{fig:tpplumian}. If we use the criterion that we need the perturbation to reenter the Hubble radius, we get a natural bound on the duration of inflation which is of direct astrophysical relevance. This portion of the inflationary regime is marked by CE and its dynamic range can be calculated to be:
 
  \begin{equation}
\frab{a_{\rm end} }{a_i} = \left( \frac{T_0 L_\Lambda}{T_{\rm reheat} H_{in}^{-1}}\right)
\left( \frac{\Omega_\Lambda}{\Omega_{DM}}\right)^{1/3}=
\frab{a_F}{a_\Lambda}(2\pi T_{\rm reheat} H_{in}^{-1})^{-1} \cong 10^{25}
\end{equation} 
 for a GUTs scale inflation with $E_{GUT}=10^{14} GeV,T_{\rm reheat}=E_{GUT},\rho_{in}=E_{GUT}^4$
we have $2\pi H^{-1}_{in}T_{\rm reheat}\approx 10^5$.
If we consider a quantum gravitational, Planck scale, inflation with $2\pi H_{in}^{-1} T_{\rm reheat} = \mathcal{O} (1)$, the phases CE and DF are approximately equal. The region in the quadrilateral CEDF is the most relevant part of standard cosmology, though the evolution of the universe can extend to arbitrarily large stretches in both directions in time. 
This figure is  telling us something regarding the duality between Planck scale and Hubble scale or between the infrared and ultraviolet limits of the theory
and is closely related to the fact that $\rho_{DE}^2=\rho_{UV}\rho_{IR}$.

\subsubsection{Area scaling for energy fluctuations}\label{sec:ccnatural}

The the geometrical mean relation described above can also be presented in a different manner which allows us to learn something significant.  Consider a 3-dimensional region of size $L$ with a bounding area which scales as $L^2$. Let us assume that we
 associate with
 this region  $N$ microscopic cells of size $L_P$ 
each having a Poissonian fluctuation in energy of amount $E_P\approx 1/L_P$. Then the mean square fluctuation of energy in this region will be $(\Delta E)^2\approx NL_P^{-2}$ corresponding to the energy density
$\rho=\Delta E/L^3=\sqrt{N}/L_PL^3$. If we make the usual assumption that $N=N_{vol}\approx (L/L_P)^3$, this will give
\begin{equation}
\rho=\frac{\sqrt{N_{vol}}}{L_PL^3}=\frac{1}{L_P^4}\frab{L_P}{L}^{3/2} \quad \text {(bulk\ fluctuations)}
\end{equation} 
On the other hand, if we assume that (for reasons which are unknown), the relevant degrees of freedom scale as the surface area of the region, then $N=N_{sur}\approx (L/L_P)^2$
and the relevant energy density is
\begin{equation}
\rho=\frac{\sqrt{N_{sur}}}{L_PL^3}=\frac{1}{L_P^4}\frab{L_P}{L}^2=\frac{1}{L_P^2L^2} \quad \text {(surface\ fluctuations)}
\label{sur}
\end{equation}
If we take $L\approx L_\Lambda$, the surface fluctuations in \eq{sur} give precisely the geometric mean in \eq{geomean} which is observed. On the other hand, the bulk \textit{fluctuations} lead to an energy density which is larger by a factor 
$(L/L_P)^{1/2}$. 
 Of course, if we do not take fluctuations in energy but coherently add them, we will get $N/L_PL^3$ which is $1/L_P^4$ for the bulk and $(1/L_P)^4(L_P/L)$
for the surface. In summary, we have the hierarchy:
\begin{equation}
\rho=\frac{1}{L_P^4}\times \left[1,\frab{L_P}{L},
\frab{L_P}{L}^{3/2},
\frab{L_P}{L}^2,
\frab{L_P}{L}^4 .....\right]
\end{equation} 
in which the first one arises by coherently adding energies $(1/L_P)$ per cell with
$N_{vol}=(L/L_P)^3$ cells; the second arises from coherently adding energies $(1/L_P)$ per cell with
$N_{sur}=(L/L_P)^2$ cells; the third one is obtained by taking \textit{fluctuations} in energy and using $N_{vol}$ cells; the fourth from energy fluctuations with $N_{sur}$ cells; and finally the last one is the thermal energy of the DeSitter space if we take $L\approx L_\Lambda$ and clearly the further terms are irrelevant due to this vacuum noise. Of all these, the only viable possibility is the one that is obtained if we assume that 
\begin{itemize}
\item 
The number of active degrees of freedom in a region of size $L$ scales as $N_{sur}=(L/L_P)^2$.
\item
It is the \textit{fluctuations} in the energy that contributes to the cosmological constant \cite{cc1,cc2} and the bulk energy does not gravitate.
\end{itemize}

Recently, it has been shown --- in a series of papers, see ref.\cite{tpholo} ---  that it is possible to obtain 
classical relativity from purely thermodynamic considerations in which the surface term of the gravitational actions play a crucial role.  
The area scaling is familiar from the usual result that entropy of horizons scale as area. In fact, one can argue from general considerations that the entropy associated with \textit{any} null surface should be $(1/4)$ per unit area and will be observer dependent. Further, in cases like Schwarzschild black hole, one cannot even properly define the volume inside a horizon. A null surface, obtained as a limit of a sequence of timelike surfaces (like the $r=2M$ obtained from $r=2M+k$ surfaces with $k\to 0^+$), `loses' one dimension in the process (e.g., $r=2M+k$ is 3-dimensional and timelike for $k>0$ but is 2-dimensional and null for $k=0$) suggesting that the scaling of degrees of freedom has to change appropriately.
It is difficult to imagine that these features are unconnected and accidental and we will discuss these ideas further in Part 3.

 \subsection{Attempts on the life of $\Lambda$}
 
 Let us now turn our attention to few of the many attempts to understand the \cc\ with the choice dictated by personal bias.  A host of other approaches exist in literature, some of which can be found in \cite{catchall}.

\subsubsection{Conservative explanations of dark energy}

One of the \textit{least} esoteric ideas regarding the dark energy
is that the cosmological constant term in the FRW equations arises because we have not calculated the energy density driving the expansion of the universe correctly. The motivation for such a suggestion arises from the following fact:  The energy momentum tensor of the real universe, $T_{ab}(t,{\bf x})$ is inhomogeneous and anisotropic and will lead to a  complicated metric $g_{ab}$ if only we could solve the exact Einstein's equations
$G_{ab}[g]=\kappa T_{ab}$.
The metric describing the large scale structure of the universe should be obtained by averaging this exact solution over a large enough scale, to get $\langle g_{ab}\rangle $. But what we actually do is to average the stress tensor {\it first} to get $\langle T_{ab}\rangle $ and {\it then} solve Einstein's equations. But since $G_{ab}[g]$ is  nonlinear function of the metric, $\langle G_{ab}[g]\rangle \neq G_{ab}[\langle g\rangle ]$ and there is a discrepancy. This is most easily seen by writing
\begin{equation}
G_{ab}[\langle g\rangle ]=\kappa [\langle T_{ab}\rangle  + \kappa^{-1}(G_{ab}[\langle g\rangle ]-\langle G_{ab}[g]\rangle )]\equiv \kappa [\langle T_{ab}\rangle  + T_{ab}^{corr}]
\end{equation}
If --- based on observations --- we take the $\langle g_{ab}\rangle $ to be the standard Friedman metric, this equation shows that it has, as its  source,  \textit{two} terms:
The first is the standard average stress tensor and the second is a purely geometrical correction term
$T_{ab}^{corr}=\kappa^{-1}(G_{ab}[\langle g\rangle ]-\langle G_{ab}[g]\rangle )$ which arises because of nonlinearities in the Einstein's theory that  leads to $\langle G_{ab}[g]\rangle \neq G_{ab}[\langle g\rangle ]$. If this term can mimic the \cc\ at large scales there will be no need for dark energy and --- as a bonus --- one will solve the coincidence problem!

 The approach requires us to identify an  effective expansion factor $a_{eff}(t)$
 of an inhomogeneous universe after suitable averaging, to be sourced by terms which will lead to $\ddot a_{eff}(t)>0$ while the standard matter [with $(\rho + 3p)>0$] leads to deceleration of standard expansion factor $a(t)$. Since correct averaging of
 positive quantities  in $(\rho + 3p)$ will not lead to a negative quantity, the real hope is in defining $a_{eff}(t)$ and obtaining its dynamical equation such that
 $\ddot a_{eff}(t)>0$.  In spite of some recent attention this idea has received \cite{flucde} it is doubtful whether it will lead to the correct result when implemented properly. The reasons for my skepticism are the following:
 
 \begin{itemize}
 \item
 Any calculation in linear theory or any calculation in which special symmetries are invoked are inconclusive in settling the issue.
 The key question, of identifying a suitable analogue of expansion factor from an averaged geometry, is nontrivial and it is not clear that the answer will be unique. To stress the point by an extreme (and a bit silly) example,
 suppose we decide to call $a(t)^n$ with, say $n>2$ as the effective expansion factor $a_{eff}(t)=a(t)^n$; obviously $\ddot a_{eff}$ can be positive (`accelerating universe') even with 
 $\ddot a$ being negative. So, unless one has a \textit{unique} procedure to identify the expansion factor of the average universe, it is difficult to settle the issue.
 \item
 It is obvious that $T_{ab}^{corr}$ is non-zero (for an explicit example, in a completely different context of electromagnetic plane wave, see \cite{gofemw}); the question that needs to be settled is how big is it compared to $T_{ab}$. It seems unlikely that when properly done, we will get a large effect for the simple reason that the amount of mass which is contained in the nonlinear regimes in the universe today is subdominant. 
\item
 This approach is  too strongly linked to explaining the acceleration as observed by SN. Even if we decide to completely ignore all SN data, we still have reasonable evidence for dark energy and it is not clear how this approach can tackle such evidence.
 
\end{itemize}
 
Another equally conservative explanation of the cosmic acceleration will be that we are located in a large underdense region in the universe; so that, locally, the underdensity acts like negative mass and produces a repulsive force. While there has been some discussion in the literature \cite{Hbubble} as to whether observations indicate such a local `Hubble bubble', this does not seem to be a tenable explanation that one can take seriously at this stage. Again, CMBR observations indicating dark energy, for example, will not be directly affected by this feature though one does need to take into account the effect of the local void.

Finally, one should not forget that a \textit{vanishing} \cc\ is still a problem that needs an explanation. So even if all the evidence for dark energy disappears within a decade, we still need to understand why \cc\ is zero and much of what I have to say in the sequel will remain relevant. I stress this because there is a recent tendency to forget the fact that the problem of the \cc\ existed (and was recognized as a problem) long before the observational evidence for dark energy, accelerating universe etc cropped up. In this sense, \cc\ problem has an important theoretical dimension which is distinct from what has been introduced by the observational evidence for dark energy.

\subsubsection{Cosmic Lenz law}

The second simplest possibility which has been attempted in the literature several times in different guises is to try and  ''cancel out'' the \cc\ by some process,
usually quantum mechanical in origin. One can, for example, ask whether switching on a \cc\ will
lead to a vacuum polarization with an effective energy momentum tensor that will tend to cancel out the \cc.
A less subtle way of doing this is to invoke another scalar field (here we go again!) such that it can couple to 
\cc\ and reduce its effective value \cite{lenz}. Unfortunately, none of this could be made to work properly. By and large, these approaches lead to an energy density which is either $\rho_{_{\rm UV}}\propto L_P^{-4}$  or to $\rho_{_{\rm IR}}\propto L_\Lambda^{-4}$. The first one is too large while the second one is too small! 

\subsubsection{Unimodular Gravity}

One possible way of addressing the issue of \cc\ is to simply eliminate from the gravitational theory those modes which couple to cosmological constant. If, for example, we have a theory in which the source in Eq. (\ref{nextnine}) is
$(\rho +p)$ rather than $(\rho +3p)$, then \cc\ will not couple to gravity at all.  Unfortunately
it is not possible to develop a covariant theory of gravity using $(\rho +p)$ as the source. But we can probably gain some insight from the following considerations. Any metric $g_{ab}$ can be expressed in the form $g_{ab}=f^2(x)q_{ab}$ such that
${\rm det}\, q=1$ so that ${\rm det}\, g=f^4$. From the action functional for gravity
\begin{equation}
A=\frac{1}{16\pi G}\int \sqrt{-g}\, d^4x\,  (R -2\Lambda)
=\frac{1}{16\pi G}\int \sqrt{-g} \, d^4x\,  R -\frac{\Lambda}{8\pi G}\int d^4x f^4(x)
\label{oneone}
\end{equation}
it is obvious that the \cc\ couples {\it only} to the conformal factor $f$. So if we consider a theory of gravity in which $f^4=\sqrt{-g}$ is kept constant and only $q_{ab}$ is varied, then such a model will be oblivious of
direct coupling to \cc. If the action (without the $\Lambda$ term) is varied, keeping ${\rm det}\, g=-1$, say, then one is lead to a {\it unimodular theory of gravity} that has  the equations of motion 
\begin{equation}
R_{ab}-(1/4)g_{ab}R=\kappa(T_{ab}-(1/4)g_{ab}T)
\label{tracefree}
\end{equation} 
with zero trace on both sides. Using the Bianchi identity, it is now easy to show that this is equivalent to the usual  theory with an {\it  arbitrary} \cc. That is, \cc\ arises as an undetermined integration constant in this model \cite{unimod}.

While this is all very interesting, we still need an extra physical principle to fix the value (even the sign) of \cc\ .
One possible way of doing this, suggested by \eq{oneone}, is to  interpret the $\Lambda$ term in the action as a Lagrange multiplier for the proper volume of the spacetime. Then it is reasonable to choose the \cc\ such that the total proper volume of the universe is equal to a specified number. While this will lead to a \cc\ which has the correct order of magnitude, it has an  obvious problem because  the proper four volume of the universe is infinite unless we make the spatial sections compact and restrict the range of time integration.  

Amongst all approaches, this one has some valuable ingredients for a solution to the \cc\ problem because it directly eliminates the coupling between gravity and bulk cosmological constant. But it needs to be remodelled considerably to be made viable. We will discuss in the next section how this can be done in a completely different approach to gravity which holds promise.

 \section{Gravity as an emergent phenomenon and the cosmological constant}
 
 \subsection{The necessary ingredients of a new perspective}

 In conventional approach to gravity, one derives the equations of motion
from a Lagrangian $\mathcal{L}_{\rm tot} = \mathcal{L}_{\rm grav}(g) + \mathcal{L}_{\rm matt}(g,\phi)$ where
$\mathcal{L}_{\rm grav}$ is the gravitational Lagrangian dependent on the metric and its derivative
and $\mathcal{L}_{\rm matt}$ is the matter Lagrangian which depends on both the metric and the 
matter fields, symbolically denoted as $\phi$. This total Lagrangian is integrated
over the spacetime volume with the covariant measure $\sqrt{-g} d^4x$ to obtain the 
action. In such an approach, the cosmological constant can be introduced via two different routes
which are conceptually different but operationally the same. 

First, one may decide
to take the gravitational Lagrangian to be $\mathcal{L}_{\rm grav} =(2\kappa)^{-1}(R-2\Lambda_g)$
where $\Lambda_g$ is a parameter in the  (low energy effective) action  just like
the Newtonian gravitational constant $\kappa$. This is equivalent to assuming that, even in
the absence of matter, flat spacetime is \textit{not} a solution to the field equations.
The second route through  which the cosmological constant can be introduced  is by
shifting  the matter Lagrangian by $\mathcal{L}_{\rm matt}\to \mathcal{L}_{\rm matt} - 2\lambda_m$. The equations
of motion for matter are invariant under such a transformation which implies that --- in the 
absence of gravity --- we cannot determine the value of $\lambda_m$. But such a shift is clearly
equivalent to adding a cosmological constant $2\kappa\lambda_m$ to the
$\mathcal{L}_{\rm grav}$. In general, what can be observed through gravitational interaction 
is the combination $\Lambda_{\rm tot} = \Lambda_g
+ 2\kappa\lambda_m$.  

It is clear that there are two distinct aspects to the so called cosmological
constant problem. The first question is why $\Lambda_{\rm tot} $ is very small
when expressed in natural units. Second, since $\Lambda_{\rm tot}$ could have
had two separate contributions from the gravitational and matter sectors, why
does the \textit{sum} remain so fine tuned? This question is particularly relevant because it is believed that our universe went through several phase transitions in the course of its  evolution, each of which shifts the energy momentum tensor of matter by $T^a_b\to T^a_b+L^{-4}\delta^a_b$ where $L$ is the scale characterizing the transition. For example, the GUT and Weak Interaction scales are about $L_{GUT}\approx 10^{-29}$ cm, $L_{SW}\approx 10^{-16}$ cm respectively which are tiny compared to $L_{\Lambda}$. 
Even if we take a more pragmatic approach, the observation of Casimir effect in the lab sets a bound that $L<\mathcal{O}(1)$ nanometer, leading to a $\rho$ which is about $10^{12}$ times the observed value \cite{gaurang}. Given all these, it seems reasonable to assume that gravity is quite successful in ignoring most of the energy density in the vacuum.

The transformation $\mathcal{L}\to \mathcal{L}_{\rm matt} - 2\lambda_m$  is a symmetry
of the matter sector (at least at scales below the scale of supersymmetry breaking; we shall ignore supersymmetry in what follows). The matter equations of motion do not care about constant $\lambda_m$.
In the conventional approach, gravity breaks this
symmetry.  \textit{This is the root cause of the so called cosmological constant problem.}
As long as
gravitational field equations are of the form $E_{ab} = \kappa T_{ab}$ where $E_{ab}$ is some geometrical quantity (which is $G_{ab}$ in Einstein's theory) the theory
cannot be invariant under the shifts of the form $T^a_b \to T^a_b +\rho \delta^a_b$.
Since such shifts are allowed by the matter sector, it is very difficult to imagine a definitive  solution
to cosmological constant problem within the conventional approach to gravity.

If metric represents the gravitational degree of freedom that is varied in the action and we demand full general covariance (unlike in the unimodular theory of gravity), we cannot  avoid $\mathcal{L}_{matter}\sqrt{-g}$ coupling and cannot obtain  of the equations of motion which are invariant under the shift
$T_{ab}\to T_{ab}+\Lambda g_{ab}$.
 Clearly a new, drastically different, approach to gravity is required.
 
 Even if we manage to obtain a theory in which gravitational action is invariant under the shift $T_{ab}\to T_{ab}+\Lambda g_{ab}$,  we would have only   succeeded in making gravity is decouple from the bulk vacuum energy. While this is considerable progress, there still remains the second issue of explaining the observed value of the \cc. Once the bulk value of the \cc\ (or vacuum energy) decouples from gravity, classical gravity becomes immune to \cc; that is, the bulk classical \cc\ can be gauged away.
Any observed value of the \cc\ has to be necessarily a quantum phenomenon arising as a relic of microscopic spacetime fluctuations. This is a nontrivial issue to address at least for two reasons: First, even the structure of \textit{matter} vacuum in the presence of nontrivial metric is far from simple; for example, it is 
 well known that the vacuum state  depends on the class of observers we are considering \cite{probe} and it is not clear whether this aspect has any fundamental significance. Second, and more important, we have no clue as to what is the substructure from which the spacetime arises as an excitation. The concept of gravitons is fairly useless \cite{gravitonmyth} in providing an answer to this --- inherently non-perturbative --- question. 
 
 Nevertheless, in an approach in which the surface degrees of freedom play the dominant role, rather than bulk degrees of freedom, we have a hope for obtaining the correct value for the \cc.
We have already seen that, in this case one obtains the correct result if the relevant degrees of freedom are scales as the surface area of a region rather as volume. Hence, to be considered plausible, any model should single out surface degrees of freedom in some suitable manner. 
 To summarise the above discussion, we are looking for an approach which has the following ingredients
\cite{gr06}:

\begin{itemize}
\item 
The field equations must remain invariant under the shift $\mathcal{L}_{matt}\to \mathcal{L}_{matt}+\lambda_m$ of the matter Lagrangian $\mathcal{L}_{matt}$ by a constant $\lambda_m$. That is, we need to have some kind of `gauge freedom' to absorb any $\lambda_m$. Once we have succeeded in decoupling gravity from bulk vacuum energy, we have won more than half the battle.
\item
General covariance requires using the integration measure $\sqrt{-g}d^Dx$ in actions. Since we do not want to restrict general covariance but at the same time do not want this coupling to metric tensor via $\sqrt{-g}$, it follows that \textit{metric cannot be the dynamical variable in our theory.}
\item
The discussion in section \ref{sec:ccnatural}, especially 
 Eq.~(\ref{sur}), shows that the
 relevant degrees of freedom should be linked to  surfaces in spacetime rather than
bulk regions. This is important because ---
 after we eliminate the coupling between the bulk \cc\ and gravity --- we still need to address  
 the observed value of \cc. This  is  a relic of quantum gravitational physics and should arise from  degrees of freedom which scale as the surface area.
\item
In such a approach, one should naturally obtain a theory of gravity which is more general than Einstein's theory with the latter emerging as a low energy approximation.
 
\end{itemize}

We will now describe how this can be achieved in a model in which gravity arises as an emergent phenomenon like elasticity. 
 
\subsection{Micro-structure  of the spacetime}

For reasons described above, we abandon the usual picture of treating the
metric as  the fundamental dynamical degrees of freedom of the theory and treat it as
 providing a
coarse grained description of the spacetime at macroscopic scales,
somewhat like the density of a solid ---  which has no meaning at atomic  
scales \cite{elastic}.  The unknown, microscopic degrees of freedom of
spacetime (which should be analogous to the atoms in the case of
solids), will play a role only when spacetime is probed at Planck
scales (which would be analogous to the lattice spacing of a solid
\cite{zeropoint}).  
 
Moreover, in the study of ordinary solids, one can distinguish between three levels of description. At the macroscopic level,
we have the theory of elasticity which has a life of its own and can be developed purely phenomenologically.  At the other extreme,  the microscopic description of a solid will be in terms of the statistical mechanics of a lattice of atoms and their interaction. 
Both of these are well known; but interpolating between these two limits  is the thermodynamic description of a solid at finite temperature \textit{which provides a crucial window into the existence of the corpuscular substructure of solids.} As Boltzmann taught us, heat is a form of motion
and we will not have the thermodynamic layer of description if  matter is a continuum all the way to the finest scales and atoms did not exist! \textit{The mere existence of a thermodynamic layer in the description is proof enough that there are microscopic degrees of freedom. } 

Move on from a solid to the spacetime. Again we should have three levels of description. The macroscopic level is the smooth spacetime continuum with a metric tensor $g_{ab}(x^i)$ and the  equations governing the metric have the same status as the phenomenological equations of elasticity. At the microscopic level, we  expect a quantum description in terms of the `atoms of spacetime' and some associated degrees of freedom $q_A$ which are still elusive. But what is crucial is the existence of an interpolating layer of thermal phenomenon associated with null surfaces in the spacetime.  Just as a solid  cannot exhibit thermal phenomenon if it does not have microstructure,  \textit{thermal nature of horizon, for example, cannot arise without the spacetime having a microstructure. } 

In such a picture, we normally expect the  microscopic structure of spacetime
to manifest itself only at Planck scales or near singularities of the
classical theory. However, in a manner which is not fully understood,
the horizons ---  which block information from certain classes of
observers --- link \cite{magglass} certain aspects of microscopic
physics with the bulk dynamics, just as thermodynamics can provide a
link between statistical mechanics and (zero temperature) dynamics of
a solid. The  reason is probably related to the fact that horizons
lead to infinite redshift, which probes \textit{virtual} high energy
processes; it is, however, difficult to establish this claim in
mathematical terms. 

The above paradigm, in which the gravity is an emergent phenomenon, is anchored on a fundamental relationship between the dynamics of gravity and thermodynamics of horizons \cite{paddy1} and
the following three results are strongly supportive of the above point of view:
\begin{itemize}
\item
There is a deep connection between the dynamical equations governing the metric and the thermodynamics of horizons. An explicit example  was provided in ref. \cite{paddy2}, in the case of spherically symmetric horizons in four
dimensions in which it was shown that, Einstein's equations can be interpreted as a 
thermodynamic relation $TdS=dE+PdV$ arising out of virtual
radial displacements of the horizon. Further work showed that this result is valid in \textit{all} the cases for which explicit computation can be carried out --- like in the
Friedmann models 
\cite{rongencai} as well as for rotating and time dependent horizons
in Einstein's theory \cite{dawood-sudipta-tp}. 
\item
The Hilbert Lagrangian has the structure $\mathcal{L}_{EH}\propto R\sim (\partial g)^2+ {\partial^2g}$.
In the usual approach the surface term arising from  $\mathcal{L}_{sur}\propto \partial^2g$ has to be ignored or canceled to get Einstein's equations from  $\mathcal{L}_{bulk}\propto (\partial g)^2$.  
But there is a 
peculiar (unexplained) relationship between $\mathcal{L}_{bulk}$ and $\mathcal{L}_{sur}$:
\begin{equation}
    \sqrt{-g}\mathcal{L}_{sur}=-\partial_a\left(g_{ij}
\frac{\partial \sqrt{-g}\mathcal{L}_{bulk}}{\partial(\partial_ag_{ij})}\right)
\end{equation}
This shows that the gravitational action is `holographic' with the same information being coded in both the bulk and surface terms and one of them is sufficient. One can indeed obtain Einstein's equations from an action principle which  uses \textit{only} the surface term and the  virtual displacements of horizons \cite{paris,gr06}. Since the surface term has the thermodynamic interpretation as the entropy of horizons, this establishes a direct connection between spacetime dynamics and horizon thermodynamics.
\item
Most importantly, recent work has shown that \textit{all the above results extend far beyond Einstein's theory.}
 The connection between field equations and the thermodynamic relation $TdS=dE+PdV$ 
 is not restricted to
Einstein's theory  alone, but is in fact true for the
case of the generalized, higher derivative \LL gravitational theory in
$D$ dimensions as well \cite{aseem-sudipta, cai2}. The same is true \cite{ayan} for the holographic structure of the action functional: the \LL action has the same structure and --- again --- the entropy of the horizons is related to the surface term of the action. \textit{These results show that the thermodynamic description is far more general than just Einstein's theory} and occurs in a wide class of theories in which the metric determines the structure of the light cones and null surfaces exist blocking the information.
\end{itemize}

The conventional approach to gravity fails to provide any clue on these results just as Newtonian continuum mechanics --- without corpuscular, discrete, substructure for matter --- cannot explain thermodynamic phenomena.
A natural explanation for these results requires a different approach to spacetime dynamics which I will now outline.

\subsection{Gravity from normalised vector fields}

Suppose there are certain microscopic --- as yet unknown --- degrees of freedom $q_A$, analogous to the atoms in the case of solids, described by some microscopic action functional $A_{micro}[q_A]$. In the case of a solid, the relevant long-wavelength elastic dynamics is captured by  the \textit{displacement vector field}
which occurs in the equation $x^a\to x^a+\xi^a(x)$. In the case of spacetime, we no longer want to use metric as a dynamical variable; so we  need to introduce some other degrees of freedom, analogous to $\xi^a$ in the case of elasticity, and an effective action functional based on it.  Normally, varying an action functional with respect certain degrees of freedom will lead to equations of motion determining \textit{those} degrees of freedom.  But we now make an unusual demand that varying our  action principle with respect to some (non-metric) degrees of freedom should lead to an equation of motion \textit{determining the background metric} which remains non-dynamical.

Based on the role expected to be played by surfaces in spacetime, we shall take the relevant degrees of freedom to be the normalised vector fields $n_i(x)$ in the spacetime \cite{aseementropy} with a  norm which is fixed at every event but might vary from event to event: (i.e., $n_in^i\equiv\epsilon(x)$ with $\epsilon(x)$ being a fixed function; one can choose the norm to be $0,\pm1$ at each event by our choice of the vector fields but its nature can vary from event to event.). 
That is, just as the displacement vector $\xi^a$ captures the macro-description in case of solids, the  normalised vectors (e.g., normals to surfaces) capture the essential macro-description in case of gravity in terms of an effective action $S[n^a]$. More formally, we expect the coarse graining of microscopic degrees of freedom to lead to an effective action in the  long wavelength limit:
\begin{equation}
\sum_{q_A}\exp (-A_{micro}[q_A])\longrightarrow \exp(-S[n^a])
\label{microtomac}
\end{equation} 
To proceed further we need to determine the nature of $S[n^a]$. The general form of $S[n^a]$ in such an effective description, at the quadratic order, will be:
\begin{equation}
S[n^a]=\int_\Cal{V}{d^Dx\sqrt{-g}}
    \left(4P_{ab}^{\ph{a}\ph{b}cd} \D_cn^a\D_dn^b - 
    T_{ab}n^an^b\right) \,,
\label{ent-func-2}
\end{equation}
where $P_{ab}^{\ph{a}\ph{b}cd}$ and $T_{ab}$ are two tensors and the signs, notation etc.  are chosen with hindsight. We will see that $T_{ab}$ can be identified with the matter stress-tensor.
The full action for gravity plus matter will be taken to be $S_{tot}=S[n^a]+S_{matt}$ with:
\begin{equation}
S_{tot}=\int_\Cal{V}{d^Dx\sqrt{-g}}
    \left(4P_{ab}^{\ph{a}\ph{b}cd} \D_cn^a\D_dn^b - 
    T_{ab}n^an^b\right)+\int_\Cal{V}{d^Dx\sqrt{-g}} \mathcal{L}_{matter}
\end{equation}  
with an important extra prescription: Since the gravitational sector is related to spacetime microstructure, we must \textit{first} vary the $n^a$ and \textit{then} vary the matter degrees of freedom. (In the language of path integrals, we should  integrate out the gravitational degrees of freedom $n^a$ first and use the resulting action for the matter sector.) We shall comment more fully on this point at the end of this section. 
 
We next address the crucial conceptual difference 
between the dynamics in gravity and elasticity, say, which we mentioned earlier. In the case of solids, one will write a similar functional [say, for entropy or free energy] in terms of the displacement vector $\xi^a$ and extremising it will
lead to an equation  \textit{which determines}
$\xi^a$. In the case of spacetime, we expect the variational principle to hold for   all vectors $n^a$ with constant norm and lead to a condition on  the \textit{background
metric.} Obviously, the action functional in \eq{ent-func-2} must be rather special to accomplish this and one need to impose two restrictions on the coefficients $P_{ab}^{\ph{a}\ph{b}cd}$ and $T_{ab}$ to achieve this.  First, the tensor $P_{abcd}$ should
have the algebraic symmetries similar to the Riemann tensor $R_{abcd}$
of the $D$-dimensional spacetime. Second, we need:
\begin{equation}
\D_{a}P^{abcd}=0=\D_{a}T^{ab}\,.
\label{ent-func-1}
\end{equation}
In a
complete theory, the explicit form of $P^{abcd}$ will be determined by the
long wavelength limit of the microscopic theory just as the elastic
constants can --- in principle --- be determined from the microscopic
theory of the lattice. In the absence of such a theory, we can take a cue from
the renormalization group theory and expand $P^{abcd}$
in powers of  derivatives of
the metric \cite{paris,aseementropy}. That is, we expect,
\begin{equation}
P^{abcd} (g_{ij},R_{ijkl}) = c_1\,\overset{(1)}{P}{}^{abcd} (g_{ij}) +
c_2\, \overset{(2)}{P}{}^{abcd} (g_{ij},R_{ijkl})  
+ \cdots \,,
\label{derexp}
\end{equation} 
where $c_1, c_2, \cdots$ are coupling constants and the successive terms progressively probe smaller and smaller scales.  The lowest order
term must clearly depend only on the metric with no derivatives. The next
term depends (in addition to metric) linearly on curvature tensor and the next one will be quadratic in curvature etc. It can be shown that  the m-th order term
which satisfies our constraints is \textit{unique} and is given by
\begin{equation}
\overset{(m)}{P}{}_{ab}^{\ph{a}\ph{b}cd}\propto
\AltC{c}{d}{a_3}{a_{2m}}{a}{b}{b_3}{b_{2m}}
\Riem{b_3}{b_4}{a_3}{a_3} \cdots
\Riem{b_{2m-1}}{b_{2m}}{a_{2m-1}}{a_{2m}} 
 =
\frac{\partial\LDm}{\partial R^{ab}_{{\ph{ab}cd}}}\,. 
\label{LL03}
\end{equation}
where $\AltC{c}{d}{a_3}{a_{2m}}{a}{b}{b_3}{b_{2m}}$ is the alternating tensor and the last equality shows that it
can be  expressed
as a derivative of the m th order \LL Lagrangian \cite{paris,lovelock}, given by
\begin{equation}
\Cal{L}^{(D)} = \sD{c_m\LDm}\,~;~\Cal{L}^{(D)}_m = \frac{1}{16\pi}
2^{-m} \Alt{a_1}{a_2}{a_{2m}}{b_1}{b_2}{b_{2m}}
\Riem{b_1}{b_2}{a_1}{a_2} \Riem{b_{2m-1}}{b_{2m}}{a_{2m-1}}{a_{2m}}
\,,  
\label{LL222}
\end{equation}
where the $c_m$ are arbitrary constants and \LDm\ is the $m$-th
order \LL term and we assume
$D\geq2K+1$.
(See Appendix for a brief description of \LL gravity.)
The lowest order term (which leads to Einstein's theory) is
\begin{equation}
\overset{(1)}{P}{}^{ab}_{cd}=\frac{1}{16\pi}
\frac{1}{2} \delta^{a_1a_2}_{b_1b_2} =\frac{1}{32\pi}
(\delta^a_c \delta^b_d-\delta^a_d \delta^b_c)
  \,.
\label{pforeh}
\end{equation} 
while the first order term (which gives the  Gauss-Bonnet correction) is:
\begin{equation}
\overset{(2)}{P}{}^{ab}_{cd}= \frac{1}{16\pi}
\frac{1}{2} \delta^{a_1a_2a_3a_4}_{b_1\,b_2\,b_3\,b_4}
R^{b_3b_4}_{a_3a_4} =\frac{1}{8\pi} \left(R^{ab}_{cd} -
         G^a_c\delta^b_d+ G^b_c \delta^a_d +  R^a_d \delta^b_c -
         R^b_d \delta^a_c\right) 
\label{ping}
\end{equation} 
where  the fourth order alternating  tensor is
\begin{equation}
\delta^{a_1a_2a_3a_4}_{b_1\,b_2\,b_3\,b_4} = \frac{-1}{(D-4)!}
  \epsilon^{c_1\cdots c_{D-4}a_1a_2a_3a_4}\epsilon_{c_1\cdots
  c_{D-4}b_1b_2b_3b_4} 
  \,, 
\label{alteps}
\end{equation}
The alternating tensors are totally antisymmetric in both sets of
indices and take values $+1$, $-1$ and $0$. They can be written in any
dimension as an appropriate contraction of the Levi-Civita tensor
density with itself. 
All higher orders terms are obtained in a similar manner (see Appendix).

In our paradigm based on \eqn{microtomac}, the field equations for gravity arises from extremising $S$ with respect to
variations of the  vector field $n^a$, with the constraint $\delta (n_an^a)=0$, and demanding that the
resulting condition holds for \textit{all normalized vector fields}.  Varying the normal vector field $ n^a$ after adding a
Lagrange multiplier function $\lambda(x)$ for imposing the constant norm  condition
$ n_a\delta  n^a=0$, we get 
\begin{equation}
\delta S = 2\int_\Cal{V}{d^Dx\sqrt{-g}
  \left(4P_{ab}^{\ph{a}\ph{b}cd}\D_c n^a\left(\D_d\delta n^b\right)
  - T_{ab} n^a\delta n^b - \lambda(x) g_{ab} n^a\delta n^b\right)}
  \label{ent-func-3}
\end{equation}
where we have used the symmetries of $P_{ab}^{\ph{a}\ph{b}cd}$ and
$T_{ab}$.  An integration by parts and the
condition $\D_dP_{ab}^{\ph{a}\ph{b}cd}=0$, leads to 
\begin{equation}
\delta
S=2\int_\Cal{V}{d^Dx\sqrt{-g}\left[-4P_{ab}^{\ph{a}\ph{b}cd}
  \left(\D_d\D_c n^a\right) - ( T_{ab}+ \lambda g_{ab}) n^a\right]\delta n^b}
  +8\int_{\dV}{d^{D-1}x\sqrt{h}\left[k_d
  P_{ab}^{\ph{a}\ph{b}cd}\left(\D_c n^a\right)\right]\delta n^b}
\,,
\label{ent-func-4}
\end{equation}
where $k^a$ is the $D$-vector field normal to the boundary \dV\ and
$h$ is the determinant of the intrinsic metric on \dV.  As usual, in order for
the variational principle to be well defined, we require that the
variation $\delta n^a$ of the  vector field should vanish on the
boundary. The second term in \eqn{ent-func-4} therefore vanishes, and
the condition that $S[ n^a]$ be an extremum for arbitrary variations of
$ n^a$ then becomes  
\begin{equation}
2P_{ab}^{\ph{a}\ph{b}cd}\left(\D_c\D_d-\D_d\D_c\right) n^a
-( T_{ab}+\lambda g_{ab}) n^a = 0\,,
\label{ent-func-5}
\end{equation}
where we used the antisymmetry of $P_{ab}^{\ph{a}\ph{b}cd}$ in its
upper two indices to write the first term. The definition of the
Riemann tensor in terms of the commutator of covariant derivatives
reduces the above expression to
\begin{equation}
\left(2P_b^{\ph{b}ijk}R^a_{\ph{a}ijk} -  T{}^a_b+\lambda \delta^a_b\right) n_a=0\,, 
\label{ent-func-6}
\end{equation}
and we see that the equations of motion \emph{do not contain}
derivatives with respect to $n^a$ which is, of course, the crucial point. This peculiar feature arose because
of the symmetry requirements we imposed on the tensor
$P_{ab}^{\ph{a}\ph{b}cd}$. We further require that the condition in
\eqn{ent-func-6} hold for \emph{arbitrary}  vector fields
$ n^a$. A simple argument based on local Lorentz invariance then
implies that 
\begin{equation}
2P_{b}^{\ph{b}ijk}R^{a}_{\ph{a}ijk} - T{}_b^a = -\lambda \delta{}_b^a\,,   
\label{ent-func-7}
\end{equation}
The scalar $\lambda$ is arbitrary so far
and we will now show how it can be determined in the physically
interesting cases.  To see what is involved, consider the lowest order approximation (viz. Einstein gravity) in which we take $P_{ab}^{\ph{a}\ph{b}cd}$ to be given by \eq{pforeh} so that the
 above equation reduces to: 
\begin{equation}
\frac{1}{8\pi}R^a_b - T^a_b=-\lambda\delta^a_b
\label{likeuni} 
\end{equation}
where $-\lambda$ can be an arbitrary function of the metric. Writing this
equation as $(G^a_b - 8\pi T^a_b ) = Q(g) \delta^a_b $ with $Q= -8\pi\lambda
- (1/2) R$ and using $\nabla_a G^a_b = 0 , \nabla_a T^a_b =0$ we get 
$\partial_bQ=\partial_b [ -8\pi \lambda - (1/2) R] =0$; so that $Q$ is an
undetermined integration constant, say $\Lambda$, and $\lambda$ must have
the form  $8\pi \lambda=-(1/2)R-\Lambda$. The resulting equation is
\begin{equation}
R^a_b-(1/2)R\delta^a_b=8\pi T^a_b+\Lambda\delta^a_b
\label{eom}
\end{equation}
which leads to Einstein's theory  if we identify $T_{ab}$ as the
matter energy momentum tensor using the standard Newtonian limit of the theory. \textit{Clearly, the cosmological constant
  appears as an integration constant}. The mathematical similarity with unimodular gravity is apparent; keeping the function $n_an^a=\epsilon(x)$ fixed while varying $n_a$ is equivalent to keeping $\sqrt{-g}$ fixed in unimodular gravity. Taking the trace of \eq{likeuni} will lead, for example, to \eq{tracefree} etc. But the conceptual structure is quite different and we maintain full general covariance. 

The crucial feature of the coupling between matter and gravity through $T_{ab}n^an^b$ is that, under the shift $T_{ab}\to T_{ab}+\rho_0g_{ab}$ the $\rho_0$  term in the action in \eq{ent-func-2} decouples from $n^a$ and becomes irrelevant:
\begin{equation}
\int_\Cal{V}{d^Dx\sqrt{-g}}T_{ab}n^an^b \to 
\int_\Cal{V}{d^Dx\sqrt{-g}} T_{ab}n^an^b +
\int_\Cal{V}{d^Dx\sqrt{-g}}\epsilon\rho_0
\end{equation} 
Since $\epsilon$ is not varied when $n_a$ is varied there is no coupling between $\rho_0$ and the dynamical variables $n_a$ the theory is invariant under the shift  $T_{ab}\to T_{ab}+\rho_0g_{ab}$.
 We see that the condition $n_an^a=$ constant on the  dynamical variables have led to a `gauge freedom' which allows an arbitrary integration constant to appear in the theory which can absorb the bulk cosmological constant. This was our key objective.

The same procedure works with the  more general structure in the family of theories starting with 
Einstein's GR, Gauss-Bonnet gravity etc and --- in the general case --- one obtains the field equations:
\begin{equation}
16\pi\left[ P_{b}^{\ph{b}ijk}R^{a}_{\ph{a}ijk}-\frac{1}{2}\delta^a_b\LDm\right]=
 8\pi T{}_b^a +\Lambda\delta^a_b   
\label{ent-func-71}
\end{equation}
These are identical to the field equations for \LL gravity with a cosmological constant arising as an undetermined integration constant.  To the lowest order, when we use \eqn{pforeh} for $P_{b}^{\ph{b}ijk}$, the \eqn{ent-func-71} reproduces Einstein's theory. More generally, we get Einstein's equations with
higher order corrections which are to be interpreted as emerging  
from the derivative expansion of the action functional as we probe smaller and smaller scales. 
Remarkably enough, we can derive not only Einstein's theory but even \LL theory from a dual description in terms on the normalised vectors in spacetime, \textit{without varying $g_{ab}$ in an action functional!}

To gain a bit more insight into what is going on, let us consider the on-shell value of the action functional.  
Manipulating
the covariant derivatives in \eqn{ent-func-2} and using the  field equation \eq{ent-func-71} we can write 
\begin{eqnarray}
S_{tot}|_{\rm on-shell}&=&S[n]+S_{mat}=\int_\Cal{V}{d^Dx\sqrt{-g}\left[
    4\D_d\left(P_{ab}^{\ph{a}\ph{b}cd}\left(
    \D_c n^a\right) n^b\right) 
    - 4P_{ab}^{\ph{a}\ph{b}cd}\left(\D_d\D_c n^a\right) n^b
    -T_{ab} n^a n^b \right]} +S_{mat}
\nonumber\\
&=&4\int_{\dV}{d^{D-1}x\sqrt{h}
    k_d\left(P_{ab}^{\ph{a}\ph{b}cd} n^b\D_c n^a\right)}
    +  \int_\Cal{V}
    {d^Dx\sqrt{-g}\epsilon\left(
  \LDm +\frac{\Lambda}{8\pi}  \right) 
   } + \int_\Cal{V}d^Dx\sqrt{-g}\mathcal{L}_{matter}
\label{on-shell-1}
\end{eqnarray}
where $\epsilon\equiv n_an^a$.
We see that, on shell, the only dependence on $n_a$ is through a surface term. Since the metric tensor is not dynamical, second term is irrelevant and we can now vary the matter Lagrangian with respect to matter variables to determine the behaviour of matter in a given curved spacetime, which, of course is sourced by the matter stress tensor through \eq{ent-func-71} obtained earlier.

The key new feature, which survives and depends on our original variables $n_a$ is the surface term which we shall now explore further.
Explicitly, this surface term is given by:
\begin{eqnarray}
S|_{\rm on-shell}&=&4\int_{\dV}{d^{D-1}x\sqrt{h}\,k_a\left(P^{abcd} n_c\D_b n_d\right)}\nonumber\\
&\longrightarrow&\frac{1}{8\pi}\int_{\dV}d^{D-1}x\sqrt{h}\,k_a\left( n^a\D_b n^b- n^b\D_b n^a\right)
=-\frac{1}{8\pi}\int_{\dV}d^{D-1}x\sqrt{h}\,k_i\left(n^iK+a^i\right)
\label{on-shell-2}
\end{eqnarray}
where we have manipulated a few indices using the symmetries of
$P^{abcd}$. The  expression in the second line, after the arrow, is the result for
general relativity. Note that the integrand has the
familiar structure of $k_i( n^iK+a^i)$ where $a^i= n^b\D_b n^i$ is
the acceleration associated with the vector field $ n^a$ and
$K\equiv -\D_b n^b$ is the trace of extrinsic
curvature in the standard context. 
If we restrict to a series of surfaces foliating the spacetime with $n_i$ representing their unit normals and take the boundary to be one of them, we can identify $k_i$ with $n_i$; then $a_in^i=0$ and the surface term is just
\begin{equation}
S|_{\rm on-shell}=\mp\frac{1}{8\pi}\int_{\dV}d^{D-1}x\sqrt{h} K
\label{ygh}
\end{equation} 
which is the York-Gibbons-Hawking boundary term in general relativity
\cite{gh} if we normalise $\epsilon$ to $\pm1$ depending on the nature of the surface.

It is now obvious that this term in the on-shell action will lead to the entropy of the horizons (which will be 1/4 per unit transverse area) in the case of general relativity. More formally, we treat the horizon surface as a limit of a sequence of timelike surfaces; for example, in the case of Schwarschild metric we consider surfaces with $r=2M+\delta$ with $\delta\to0$. In fact, the result is far more general. Even in the case of of a more general $P^{ab}_{cd}$ it can be shown that the on-shell value of the action reduces to \cite{aseementropy} the entropy of the horizons. The general expression is:
\begin{equation}
S|_{\Cal{H}} = \sD{4\pi m c_m \int_{\Cal{H}}{d^{D-2}x_{\perp} 
  \sqrt{\sigma}\Cal{L}^{(D-2)}_{(m-1)}}} 
 =\frac{1}{4}[{\rm Area}]_\perp +{\rm corrections}     
\label{ent-limit-2}
\end{equation} 
where $x_{\perp}$ denotes the transverse coordinates on the horizon \Cal{H},
$\sigma$ is the determinant of the intrinsic metric on \Cal{H} and we
have restored a summation over $m$ thereby giving the result for the
most general \LL case obtained as a sum of individual \LL lagrangians.  The expression in \eqn{ent-limit-2} \emph{is
  precisely the entropy of a general Killing horizon in \LL gravity}
based on the general prescription given by Wald and others
\cite{noether} and computed by several authors.  
Further, in any spacetime, 
if we take a local Rindler frame around  any event 
we will obtain an entropy for the locally defined Rindler horizon. In the case of GR, this entropy per unit transverse area is just 1/4 as expected. 

This result shows that, in the semiclassical limit in which the action can possibly be related to entropy, we reproduce the conventional entropy which scales as the area in Einstein's theory. Since the entropy counts the relevant degrees of freedom, this shows that the degrees of freedom which survives and contributes in the long wave length limit  scales as the area. The quantum fluctuations in these degrees of freedom can then lead to the correct, observed, value of the \cc.
The last aspect can be made more quantitative and we will briefly describe in the next section how this can be done.

Our action principle is somewhat peculiar compared to the usual action principles in the sense that we have varied $n_a$ and demanded that the resulting equations hold for \textit{all} vector fields of constant norm. Our action principle actually stands for an infinite number of action principles, one for each vector field of constant norm! This class of \textit{all} $n^i$ allows an effective, coarse grained, description of some (unknown) aspects of spacetime micro physics. This is why we need to first vary $n_a$, obtain the equations constraining the background metric and then use the action in \eq{on-shell-1} to obtain the equations of motion for matter.  (If, instead, we vary matter terms first the coupling $T_{ab}n^an^b$ will couple matter to $n^a$ which will remain undetermined since we have no equation for $n_a$.) Of course, in most contexts, $\nabla_a T^a_b=0$ will take care of the dynamical equations for matter and these issues are irrelevant \cite{note3}.

At this stage, it is not possible to proceed further and relate $n^i$ to some microscopic degrees of freedom $q^A$. This issue is conceptually similar to asking one to identify the atomic degrees of freedom, given the description of an elastic solid in terms of a displacement field $\xi^a$ --- which we know is impossible. However, the same analogy tells us that the relevant degree of freedom in the long wavelength limit (viz. $\xi^a$ or $n^i$) can be completely different from the microscopic degrees of freedom and it is best to proceed phenomenologically.

 \subsection{Gravity as detector of the vacuum fluctuations}

The description of gravity using the action principle given above provides a natural back drop for gauging away the bulk value of the cosmological constant since it decouples from the dynamical degrees of freedom in the theory.  Once the bulk term is eliminated, 
what is observable through gravitational effects, in the correct theory of quantum gravity, should be the \textit{fluctuations} in the vacuum energy.
These fluctuations will be non-zero if the universe has a DeSitter horizon which provides a confining 
volume. In this paradigm the
 vacuum structure can readjust  to gauge away the bulk energy density $\rho_{_{\rm UV}}\simeq L_P^{-4}$ while quantum \textit{fluctuations} can generate
the observed value $\rho_{\rm DE}$. 

The role of energy fluctuations contributing to gravity also arises, more formally, when we study the question of \emph{detecting} the energy
density using gravitational field as a probe.
 Recall that an Unruh-DeWitt detector with a local coupling $\mathcal{L}_I=M(\tau)\phi[x(\tau)]$ to the {\it field} $\phi$
actually responds to $\langle 0|\phi(x)\phi(y)|0\rangle$ rather than to the field itself \cite{probe}. Similarly, one can use the gravitational field as a natural ``detector" of energy momentum tensor $T_{ab}$ with the standard coupling $L=\kappa h_{ab}T^{ab}$. Such a model was analyzed in detail in ref.~\cite{tptptmunu} and it was shown that the gravitational field responds to the two point function $\langle 0|T_{ab}(x)T_{cd}(y)|0\rangle $. In fact, it is essentially this fluctuations in the energy density which is computed in the inflationary models \cite{inflation} as the  {\it source} for gravitational field, as stressed in
ref.~\cite{tplp}. All these suggest treating the energy fluctuations as the physical quantity ``detected" by gravity, when
one  incorporates quantum effects.  

If the \cc\ arises due to the fluctuations in the energy density of the vacuum, then one needs to understand the structure of the quantum gravitational vacuum at cosmological scales. Quantum theory, especially the paradigm of renormalization group has taught us that the  concept of the vacuum
state  depends on the scale at which it is probed. The vacuum state which we use to study the
lattice vibrations in a solid, say, is not the same as vacuum state of the QED
 and it is not appropriate to ask questions about the vacuum without specifying the scale. 
 If the spacetime has a cosmological horizon which blocks information, the natural scale is provided by the size of the horizon,  $L_\Lambda$, and we should use observables defined within the accessible region. 
The operator $H(<L_\Lambda)$, corresponding to the total energy  inside
a region bounded by a cosmological horizon, will exhibit fluctuations  $\Delta E$ since vacuum state is not an eigenstate of 
{\it this} operator. The corresponding  fluctuations in the energy density, $\Delta\rho\propto (\Delta E)/L_\Lambda^3=f(L_P,L_\Lambda)$ will now depend on both the ultraviolet cutoff  $L_P$ as well as $L_\Lambda$.  
 To obtain
 $\Delta \rho_{\rm vac} \propto \Delta E/L_\Lambda^3$ which scales as $(L_P L_\Lambda)^{-2}$
 we need to have $(\Delta E)^2\propto L_P^{-4} L_\Lambda^2$; that is, the square of the energy fluctuations
 should scale as the surface area of the bounding surface which is provided by the  cosmic horizon.  
 Remarkably enough, a rigorous calculation \cite{cc2} of the dispersion in the energy shows that
 for $L_\Lambda \gg L_P$, the final result indeed has  the scaling 
 \begin{equation}
 (\Delta E )^2 = c_1 \frac{L_\Lambda^2}{L_P^4} 
 \label{deltae}
 \end{equation}
 where the constant $c_1$ depends on the manner in which ultra violet cutoff is imposed.
 Similar calculations have been done (with a completely different motivation, in the context of 
 entanglement entropy)
 by several people and it is known that the area scaling  found in Eq.~(\ref{deltae}), proportional to $
L_\Lambda^2$, is a generic feature \cite{area}.
For a simple exponential UV-cutoff, $c_1 = (1/30\pi^2)$ but  cannot be computed
 reliably without knowing the full theory.
  We thus find that the fluctuations in the energy density of the vacuum in a sphere of radius $L_\Lambda$ 
 is given by 
 \begin{equation}
 \Delta \rho_{\rm vac}  = \frac{\Delta E}{L_\Lambda^3} \propto L_P^{-2}L_\Lambda^{-2} \propto \frac{H_\Lambda^2}{G}
 \label{final}
 \end{equation}
 The numerical coefficient will depend on $c_1$ as well as the precise nature of infrared cutoff 
 radius;
 but it is a fact of life that a fluctuation of magnitude $\Delta\rho_{vac}\simeq H_\Lambda^2/G$ will exist in the
energy density inside a sphere of radius $H_\Lambda^{-1}$ if Planck length is the UV cut off. 
On the other hand, since observations suggest that there is a $\rho_{vac}$ of similar magnitude in the universe it seems 
natural to identify the two. Our approach explains why there is a \textit{surviving} cosmological constant which satisfies 
$\rho_{_{\rm DE}}=\sqrt{\rho_{_{\rm IR}}\rho_{_{\rm UV}}}$.
  
We stress that the computation of energy fluctuations is completely meaningless in the conventional models of gravity in which the metric couples to the bulk energy density. Once a UV cut-off at Planck scale is imposed, one will always get a bulk contribution $\rho_{UV}\approx L_P^{-4}$ with  the usual problems. It is only because we have a way of decoupling the bulk term  from contributing to the dynamical equations that, we have a right to look at the subdominant term $L_P^{-4}(L_P/L_\Lambda)^2$. Approaches in which the sub-dominant term is introduced by an ad hoc manner are technically flawed since the bulk term cannot be ignored in these usual approaches to gravity.
Getting the correct value of the cosmological constant from the energy fluctuations is not as difficult as understanding why the bulk value  (which is larger
by $10^{120}$!) can be ignored. Our approach provides a natural backdrop for 
ignoring the bulk term --- and as a bonus --- we get the right value for the cosmological 
constant from the fluctuations. It  is small because it is a purely quantum effect.

\section{Conclusions}

It is obvious that  the existence of a component with negative pressure constitutes a major challenge in theoretical physics.
 The simplest choice for this component is the cosmological constant; other models based on scalar fields [as well as those based on branes etc. which I  have not discussed] do not alleviate the difficulties faced by \cc\  and --- in fact --- makes them worse. 
 The key point I want to stress is that the cosmological constant
  is most likely to be a low energy relic of a quantum gravitational effect or principle and its explanation will require a radical shift in our current paradigm.
  
I have tried to advertise a new approach to gravity as a possible broad paradigm
to understand the cosmological constant. On the negative side, there are some very obvious difficulties with the ideas that I have outlined. The most serious objections are the following:
\begin{itemize}
\item
The normalised vectors $n_i$ were introduced in a totally ad hoc manner and does not relate to anything we know about gravity and hence the motivation for the condition the $n_in^i=$ constant  is unclear. The unusual nature of this variable and the action $S[n_a]$ makes it difficult to construct a quantum theory via path integrals.
\item
While we have fairly attractive scheme to eliminate the bulk \cc\ term, the arguments given in the last section to obtain the observed value is, at best, tentative. The area scaling for surviving degrees of freedom emerges naturally but it is unclear how to connect up the energy fluctuations in these degrees of freedom to the source of gravity.
\end{itemize}
  
Against this, one should compare the attractive features of the approach  in a broader context.  
 The conceptual
  basis for this approach rests on the following logical ingredients.
  \begin{enumerate}
\item  
It is impossible to solve the \cc problem unless the gravitational sector of the theory is invariant under the shift $T_{ab}\to T_{ab}+\lambda_mg_{ab}$. Any approach which does not address this issue cannot provide a comprehensive solution to the \cc\ problem.
\item  
General covariance requires us to use the measure $\sqrt{-g}d^Dx$ in D-dimensions
in the action. This will couple the metric (through its determinant) to the matter sector. Hence, as long as we insist on metric as the fundamental variable describing gravity, one cannot address the issue in (1) above.  So we need to introduce some other degrees of freedom and an effective action which, however, is capable of constraining the background metric.
\item
We found an action principle, based on the  normalised vector fields in spacetime, that satisfies all these criteria mentioned above. The new action does not couple to the bulk energy density and maintains invariance under the shift $T_{ab}\to T_{ab}+\lambda_mg_{ab}$. What is more, the on shell value of the action is related to the entropy of horizons showing the relevant degrees of freedom scales as the area of the bounding surface.
\item
Since our formalism ensures that the bulk energy density does not contribute to gravity --- and only because of that --- it makes sense to compute the next order correction due to fluctuations in the energy density. This is impossible to do rigorously with the machinery available but a plausible case can be made as how this will lead to the correct, observed, value of the \cc.

\item  
In the long wavelength limit, the relevant physics is  captured in terms of an effective theory related to the degrees of freedom contained in the fluctuations of the normalised vectors. 
The  
resulting theory  is far more general than Einstein gravity since the thermodynamic interpretations should transcend classical considerations and incorporate some of the microscopic corrections. 
Einstein's equations provide the lowest
order description of the dynamics and \textit{calculable}, higher order,
corrections arise as we probe smaller scales.    
 The mechanism for ignoring the bulk \cc\ is likely to survive quantum gravitational corrections which are likely to bring in additional, higher derivative, terms to the action.
 
\end{enumerate}

Taking stock, I strongly believe there is no way out of the points mentioned in (1) and (2) above and a tenable description of gravity must be based on variables other than the metric. Such a theory is very likely to have most of the ingredients I have outlined here.

\section{Acknowledgements}
I thank A.Paranjape and K.Subramanian for useful comments on first draft of the review.
 
\section*{Appendix: A primer on \LL gravity}
The \LL Lagrangian is a specific example from a general
class of Lagrangians which describes a (possibly semiclassical) theory
of gravity and are given by    
\begin{equation}
\Cal{L}=Q_a^{\ph{a}bcd}R^a_{\ph{a}bcd}\,,
\label{LL1}
\end{equation}
where $Q_a^{\ph{a}bcd}$ is the most general fourth rank tensor sharing
the algebraic symmetries of the Riemann tensor $R^a_{\ph{a}bcd}$ and further
satisfying the criterion $\D_bQ_a^{\ph{a}bcd}=0$ (Several general properties of this class of Lagrangians are discussed in Ref. \cite{ayan}). 
 It can
be shown  that (see e.g., \cite{ayan}) the equations of motion 
for a general theory of gravity derived from the Lagrangian in
\eqn{LL1} using the standard variational principle with $g^{ab}$ as
the dynamical variables, are given by  
\begin{equation}
E_{ab}=\frac{1}{2}T_{ab} ~~;~~
  E_{ab}\equiv\frac{1}{\sqrt{-g}}\frac{\partial}{\partial
  g^{ab}}\left(\sqrt{-g}\Cal{L}\right) -2\D^m\D^nP_{amnb}\,. 
\label{LL4}
\end{equation}
Here $T_{ab}$ is the energy-momentum tensor for the matter fields. The 
tensor $P_{abcd}$ defined through
$P_a^{\ph{a}bcd}\equiv(\partial\Cal{L}/\partial R^a_{\ph{a}bcd})$.
The partial derivatives are  taken treating $g^{ab}$, 
$\Gamma^a_{\ph{a}bc}$ and $R^a_{\ph{a}bcd}$ as independent
quantities.
 
The $D$-dimensional
\LL Lagrangian is given by \cite{lovelock} a polynomial in the
curvature tensor: 
\begin{equation}
\Cal{L}^{(D)} = \sD{c_m\LDm}\,~;~\Cal{L}^{(D)}_m = \frac{1}{16\pi}
2^{-m} \Alt{a_1}{a_2}{a_{2m}}{b_1}{b_2}{b_{2m}}
\Riem{b_1}{b_2}{a_1}{a_2} \Riem{b_{2m-1}}{b_{2m}}{a_{2m-1}}{a_{2m}}
\,,  
\label{LL2}
\end{equation}
where the $c_m$ are arbitrary constants and \LDm\ is the $m$-th
order \LL term. Here the generalised alternating 
tensor $\delta^{\cdots}_{\cdots}$  is the natural extension of the one
defined in \eqn{alteps} for $2m$ indices, and we assume
$D\geq2K+1$. The $m$-th order \LL term $\Cal{L}^{(D)}_m$ given  
in \eqn{LL2} is a homogeneous function of the Riemann tensor of degree
$m$. For each such term, the tensor $Q_a^{\ph{a}bcd}$ defined in
\eqn{LL1} carries a label $m$ and becomes 
\begin{equation}
{}^{(m)}Q_{ab}^{\ph{a}\ph{b}cd}=  \frac{1}{16\pi}2^{-m}
\AltC{c}{d}{a_3}{a_{2m}}{a}{b}{b_3}{b_{2m}}
\Riem{b_3}{b_4}{a_3}{a_3}\cdots\Riem{b_{2m-1}}{b_{2m}}{a_{2m-1}}{a_{2m}}
\,.
\label{LL3}
\end{equation}
The full tensor $Q_{ab}^{\ph{a}\ph{b}cd}$ is a linear combination of
the ${}^{(m)}Q_{ab}^{\ph{a}\ph{b}cd}$ with the coefficients $c_m$.
Einstein's GR is a special case of \LL gravity in which only the
coefficient $c_1$ is non-zero. Since the tensors
${}^{(m)}Q_{ab}^{\ph{a}\ph{b}cd}$ appear linearly in the \LL
Lagrangian and consequently in all other tensors constructed from it,
for most applications it is sufficient to concentrate on the case where a single coefficient
$c_m$ is non-zero. All the results that follow can be easily extended
to the case where more than one of the $c_m$ are non-zero, by taking
suitable linear combinations of the tensors involved.

 For the $m$-th order \LL Lagrangian $\LDm$, since
$P^{abcd}$ is divergence-free, the expression for the tensor $E_{ab}$
in \eqn{LL4} becomes 
\begin{equation}
E_{ab}=\frac{\partial\LDm}{\partial g^{ab}}-\frac{1}{2}\LDm 
g_{ab}\,, 
\label{LL5}
\end{equation}
where we have used the relation $\partial(\sqrt{-g})/\partial
g^{ab}=-(1/2)\sqrt{-g}g_{ab}$. The first term in the expression for
$E_{ab}$ in \eqn{LL5} can be simplified to give 
\begin{equation}
\frac{\partial\LDm}{\partial
  g^{ab}}=mQ_{a}^{\ph{a}ijk}R_{bijk}= P_a^{\ph{a}ijk}R_{bijk} \,,  
\label{LL6}
\end{equation}
where the expressions in \eqn{LL6} can be verified by direct
computation, or by noting that \LDm\ is a homogeneous function 
of the Riemann tensor $R^a_{\ph{a}bcd}$ of degree $m$. To summarize,
the \LL field equations are given by
\begin{equation}
16\pi\left[ P_{b}^{\ph{b}ijk}R^{a}_{\ph{a}ijk}-\frac{1}{2}\delta^a_b\LDm\right]=
 8\pi T{}_b^a,   
\label{ent-func-711}
\end{equation}
where we have included a possible cosmological constant in the defintion of $T^a_b$. Taking the trace of this equation, we find that that $\LDm=(2m-D)^{-1}T$. In other words, the on-shell value of the Lagrangian is proportional to the trace of the stress tensor in all \LL theories, just like in GR. 
.


\begin{thebibliography}{000}

\bibitem{cmbr}
P.  de Bernardis  et al., (2000), Nature \textbf{404}, 955; 
A.   Balbi  et al., (2000), Ap.J., \textbf{545}, L1; 
S. Hanany et al., (2000), Ap.J., \textbf{545}, L5; 
T.J.   Pearson  et al., Astrophys.J., \textbf{591} (2003) 556;
C.L. Bennett et al, Astrophys.~J.~Suppl. ,{\bf 148}, 1 (2003);
D.~N.~Spergel et al., ApJS, {\bf 148}, 175 (2003);
B. S. Mason et al., Astrophys.J. ,\textbf{591 }(2003) 540;
D. N. Spergel et al., astro-ph/0603449.

\bibitem{baryon} 
W.J. Percival  et al., Mon.Not.Roy.Astron.Soc. \textbf{337}, 1068 (2002);  MNRAS {\bf  327}, 1297 (2001);
X.  Wang et al., (2002), Phys. Rev. {\bf  D 65}, 123001;
T. Padmanabhan  and Shiv Sethi, Ap. J, (2001), {\bf 555}, 125 [astro-ph/0010309].
For a review of BBN, see  S.Sarkar, Rept.Prog.Phys., (1996), \textbf{59}, 1493.

\bibitem{h} 
W. Freedman  {\it et al.}, (2001),  Astrophysical Journal,\textbf{ 553},  47;
J.R. Mould  etal., Astrophys. J., (2000), {\bf 529},  786.

  
\bibitem{adcos} For recent reviews of cosmological paradigm, see, e.g.,  
T. Padmanabhan, \textit{Advanced topics in Cosmology: A pedagogical introduction},  AIP Conference Proceedings, 2006,  843,111-166, [astro-ph/0602117];
T. Padmanabhan \textit{Understanding Our Universe: Current Status and Open Issues} in, `100 Years of Relativity - Space-time Structure: Einstein and Beyond', A.Ashtekar (Editor), World Scientific (Singapore, 2005) pp 175-204; [gr-qc/0503107]


\bibitem{inflation} 
D.~Kazanas, \textit{Ap. J. Letts. } {\bf 241}, 59 (1980); 
A.~A.~Starobinsky, \textit{JETP\ Lett.} {\bf 30}, 682  (1979); 
                    \textit{Phys. Lett. } {\bf B 91}, 99  (1980);
A.~H.~Guth, \textit{Phys. Rev.  } {\bf D 23}, 347 (1981); 
A.~D.~Linde, \textit{Phys. Lett.}  {\bf B 108}, 389  (1982); 
A.~Albrecht, P.~J.~Steinhardt, \textit{Phys. Rev. Lett.} {\bf 48}, 1220  (1982); 
for a review, see e.g.,J.~V.~Narlikar and T.~Padmanabhan, \textit{Ann. Rev. Astron. Astrophys.}  {\bf 29}, 325 (1991);
L.~Alabidi and D.~H.~Lyth, astro-ph/0510441.

\bibitem{genofpert} 
S.~W.~Hawking,  \textit{Phys. Lett.} {\bf B 115}, 295   (1982); 
A.~A.~Starobinsky,\textit{ Phys. Lett.}   {\bf B 117}, 175 (1982); 
A.~H.~Guth, S.-Y.~Pi, \textit{Phys. Rev. Lett.} {\bf 49}, 1110  (1982); 
J.~M.~Bardeen et al., \textit{Phys. Rev. } {\bf D 28}, 679  (1983);
L.~F.~Abbott, M.~B.~Wise, \textit{Nucl. Phys.} {\bf B 244}, 541   (1984).
For a recent discussion with detailed set of references, see  
L. Sriramkumar,  T. Padmanabhan, Phys. Rev., \textbf{D 71}, 103512 (2005) [gr-qc/0408034].  
 

\bibitem{tplp}
T.  Padmanabhan,  \textit{Phys. Rev. Letts.} , (1988), \textbf{60} , 2229;
T.  Padmanabhan, T.R. Seshadri and T.P. Singh,  \textit{Phys.Rev.}  \textbf{D 39}, 2100 (1989).
 

\bibitem{cobeanaly} 
G.F. Smoot. et al., \textit{Ap.J.} \textbf{396}, L1 (1992); 
T. Padmanabhan, D. Narasimha, \textit{MNRAS} \textbf{259}, 41P (1992); 
G. Efstathiou et al.,\textit{ MNRAS},  \textbf{258}, 1  (1992).


\bibitem{nlapprox} 
Ya.B. Zeldovich, \textit{Astron.Astrophys.} \textbf{5}, 84 (1970); 
Gurbatov, S. N. et al, \textit{MNRAS} \textbf{236}, 385 (1989);  
T.G. Brainerd et al., \textit{Astrophys.J.} \textbf{418}, 570 (1993);
J.S. Bagla, T.Padmanabhan, \textit{MNRAS} \textbf{ 266}, 227 (1994), [gr-qc/9304021];
                 \textit{ MNRAS}, \textbf{ 286}, 1023 (1997), [astro-ph/9605202];
T.Padmanabhan, S.Engineer,   \textit{Ap. J.}   \textbf{493}, 509 (1998), [astro-ph/9704224];  
S. Engineer  et.al., \textit{ MNRAS}   \textbf{314}, 279 (2000), [astro-ph/9812452]; 
This is essentially an example of statistical mechanics
of self gravitating systems; see e.g., T.Padmanabhan,  \textit{Phys. Rept.}  \textbf{ 188}, 285 (1990); \textit{Astrophys. Jour. Supp.}, \textbf{ 71} , 651 (1989); T.Tatekawa, [astro-ph/0412025].
  

  
\bibitem{nsr} 
A.~J.~S. Hamilton et al., \textit{ Ap. J.} \textbf{374}, L1 (1991); 
R. Nityananda, T. Padmanabhan, \textit{MNRAS} \textbf{271}, 976 (1994), [gr-qc/9304022]; 
T. Padmanabhan, \textit{MNRAS} \textbf{278}, L29 (1996), [astro-ph/9508124];astro-ph/0512077;
T.Padmanabhan et al.,\textit{ Ap. J.} \textbf{466}, 604 (1996), [astro-ph/9506051];  D. Munshi et al., \textit{ MNRAS},  \textbf{290}, 193 (1997), [astro-ph/9606170];
J.~S. Bagla, et.al.,   \textit{Ap.J.} \textbf{495}, 25 (1998), [astro-ph/9707330];
N.Kanekar et al., \textit{MNRAS }, \textbf{ 324}, 988 (2001), [astro-ph/0101562];
 T.Padmanabhan, S. Ray,  Mon.Not.Roy.Astron.Soc.Letters, \textbf{372}: L53-L57 (2006)[astro-ph/0511596]

 
\bibitem{baryonsimulations} 
For a pedagogical description, see 
J.S. Bagla,  astro-ph/0411043; 
J.S. Bagla, T. Padmanabhan, \textit{Pramana} \textbf{49}, 161-192 (1997), [astro-ph/0411730].


\bibitem{earlyde} 
G.~Efstathiou et al.,  Nature,  (1990), \textbf{348}, 705;
J.~P.~Ostriker and P.~J.~Steinhardt, Nature,  (1995), {\bf 377}, 600;
J.~S.~Bagla, T.~Padmanabhan and J.~V.~Narlikar, Comments   on Astrophysics,  (1996), \textbf{18}, 275 [astro-ph/9511102].
 

 
\bibitem{sn} 
S.J. Perlmutter et al., Astrophys. J. (1999) \textbf{517},565;
A.G. Reiss et al., Astron. J. (1998), \textbf{116},1009;
J.~L.~Tonry et al., ApJ,  (2003), {\bf 594}, 1;
B.~J.~Barris, Astrophys.J., \textbf{602} (2004), 571;
A.~G.Reiss et al., Astrophys.J. \textbf{607}, (2004), 665.

\bibitem{snls}
 P. Astier et al.,Astron.Astrophys., \textbf{447}, 31 (2006) [astro-ph/0510447].

 
\bibitem{tptirthsn1} 
T.~Padmanabhan,T.~Roy~Choudhury, \textit{MNRAS} {\bf 344}, 823 (2003) [astro-ph/0212573];
T.~Roy~Choudhury, T.~Padmanabhan, \textit{Astron.Astrophys.} \textbf{429}, 807 (2005), [astro-ph/0311622];
H.K.Jassal et al., \textit{ Phys.Rev.} \textbf{D 72}, 103503 (2005) [astro-ph/0506748];
S. Nesseris, L.Perivolaropoulos, JCAP \textbf{0702},025 (2007).

\bibitem{sndataanalysis} 
A sample of recent (2007) work in SN data and related topics is:
Y.Gong, A.Wang,arXiv:0705.0996;
R. Lazkoz, et al.,arXiv:0704.2606;
Y.Gong et al.,astro-ph/0703583;
R.Rosenfeld, Phys.Rev.\textbf{D75},083509 (2007);
D. Rapetti et al., Mon.Not.Roy.Astron.Soc.\textbf{375}, 1510 (2007);
A.A. Sen, R.J. Scherrer, astro-ph/0703416;
A. Shafieloo, astro-ph/0703034;
C.Cattoen, M.Visser, gr-qc/0703122.

\bibitem{cc} 
T.~Padmanabhan, \textit{Phys. Rept.} {\bf 380}, 235 (2003) [hep-th/0212290]; 
               \textit{Current Science}, \textbf{88},1057, (2005) [astro-ph/0411044];
S.Nobbenhuis, gr-qc/0609011;
J.S. Alcaniz,astro-ph/0608631;
S. Hannestad, Int.J.Mod.Phys., \textbf{A21},1938 (2006);
L. Perivolaropoulos, AIP Conf.Proc.,\textbf{848},698 (2006);
E. J. Copeland et al.,Int.J.Mod.Phys., \textbf{D15}, 1753 (2006);
A.D. Dolgov, hep-ph/0606230;
P.~J.~E.~Peebles and B.~Ratra, \textit{Rev. Mod. Phys.}  {\bf 75}, 559 (2003);
V. Sahni, A. Starobinsky, Int.J.Mod.Phys.\textbf{D9}, 373 (2000).


\bibitem{phiindustry} 
A  sample of recent ($\gtrsim2003$) papers covering time varying $w$ in different guises are:
A. Fuzfa, J.-M. Alimi,astro-ph/0611284;
J.A.S. Lima et al.,astro-ph/0611007;
A. Das, et al.,gr-qc/0610097;
J.C. Fabris et al., gr-qc/0609017;
R. Gannouji et al.,astro-ph/0606287;
C. J. Gao et al.,astro-ph/0605682;
J. Grande et al., gr-qc/0604057;
S. Nojiri, S.D. Odintsov.  hep-th/0601213;hep-th/0606025;
G. Panotopoulos, astro-ph/0606249;
S. Carneiro et al., astro-ph/0605607;
V. B. Johri, P.K. Rath, astro-ph/0603786;
M. Wang. hep-th/0601189;
H. Wei et al., \textit{Phys.Rev}. \textbf{D 72}, 123507 (2005);
S. Capozziello, astro-ph/0508350;
D. Polarski, A. Ranquet, \textit{Phys.Lett}. \textbf{B 627}, 1 (2005);
A. A. Andrianov et al., \textit{Phys.Rev}. \textbf{D 72}, 043531 (2005);
H.Stefancic, astro-ph/0504518;
J.Sola, H.Stefancic, \textit{Mod.Phys.Lett}. \textbf{A21} (2006) 479; \textit{Phys.Lett.} \textbf{B624} (2005) 147;
M. Sahlen et al., astro-ph/0506696;
Zhuo-Yi Huang et al., astro-ph/0511745;
Z. K. Guo, N. Ohta and Y. Z. Zhang, astro-ph/0505253;
S. Nojiri, S. D. Odintsov, hep-th/0506212; hep-th/0408170;
W. Godlowski et al., \textit{Ap.J}., \textbf{605}, 599 (2004); astro-ph/0604327;
V.~F.~Cardone et al., Phys.Rev. \textbf{D69}, (2004), 083517;
I.P. Neupane, \textit{Class.Quant.Grav}. \textbf{21}, 4383 (2004);
M.C. Bento et al., astro-ph/0407239;
A. DeBenedictis et al., gr-qc/0402047;
M. Axenides and K.Dimopoulos, hep-ph/0401238;
M. D. Maia et al., astro-ph/0403072;
J. S. Alcaniz,  astro-ph/0312424;
Xin-Zhou Li et al., \textit{Int.J.Mod.Phys}. \textbf{A18}, 5921 (2003);
P.~J.~Steinhardt, \textit{Phil.\ Trans.\ Roy.\ Soc.\ Lond.}  {\bf A 361}, 2497 (2003);
S.~Sen and T.~R.~Seshadri, \textit{Int. J. Mod. Phys.}  {\bf D 12}, 445 (2003);
P.~F.~Gonzalez-Diaz, Phys.\ Lett.\ B {\bf 562}, 1 (2003).

\bibitem{kessence}
A small sample of recent ($\gtrsim2003$) papers are
R. de Putter, Eric V. Linder, arXiv:0705.0400;
M.S. Movahed et al.,astro-ph/0701339;
 Hui Li et al., astro-ph/0601007;
L.~P.~Chimento and A.~Feinstein, \textit{Mod. Phys. Lett.}  {\bf A 19}, 761 (2004);
R.~J.~Scherrer,  \textit{Phys. Rev. Lett.} \textbf{93} ) 011301  (2004;
 P. F. Gonzalez-Diaz,hep-th/0408225;
  L. P. Chimento, \textit{Phys.Rev.} \textbf{D69}, 123517, (2004);
  O.Bertolami, astro-ph/0403310;
  R.Lazkoz, gr-qc/0410019;
J.S. Alcaniz, J.A.S. Lima, astro-ph/0308465;
M.~Malquarti et al., \textit{Phys. Rev.}  {\bf D 67}, 123503 (2003).

\bibitem{tptachyon}
T.~Padmanabhan, \textit{Phys. Rev. } {\bf D 66}, 021301 (2002), [hep-th/0204150];
T.~Padmanabhan and T.~R.~Choudhury, \textit{Phys. Rev. } {\bf D 66}, 081301 (2002) [hep-th/0205055];
J.~S.~Bagla, et al., \textit{Phys.   Rev.}  {\bf D  67}, 063504 (2003) [astro-ph/0212198].

\bibitem{tachyon} 
For a varied sample, see
Y. Shao et al., gr-qc/0703112;
V. Zamarias, hep-th/0610063;
J. Ren et al., astro-ph/0610266;
W. Fang et al., hep-th/0606033;
G. Calcagni, Andrew R. Liddle, astro-ph/0606003;
A.A. Sen, gr-qc/0604050;
H. Singh, hep-th/0608032; hep-th/0505012;
 A. Das et al., \textit{Phys.Rev}. \textbf{D 72}, 043528 (2005);
 I.Ya. Aref'eva et al., astro-ph/0505605; astro-ph/0410443;
 J.~M.~Aguirregabiria and R.~Lazkoz, hep-th/0402190;
C. Kim et al., hep-th/0404242;
R. Herrera et al.,astro-ph/0404086;
A. Ghodsi and A.E.Mosaffa,hep-th/0408015;
D.J. Liu and X.Z.Li, astro-ph/0402063;
V.~Gorini et al.,  \textit{Phys.Rev.} \textbf{D 69}  123512 (2004);
M. Sami et al., \textit{Pramana} \textbf{62},  765 (2004);
D.A. Steer, \textit{Phys.Rev}. \textbf{D70}, 043527 (2004);
  L.Raul W. Abramo, F. Finelli, \textit{Phys.Lett}. \textbf{B 575}, 165 (2003);
 L. Frederic, A. W. Peet, \textit{JHEP} \textbf{0304}, 048 (2003);
 M. Sami, \textit{Mod.Phys.Lett}. \textbf{A 18}, 691 (2003);
C.~J.~Kim et al.,  \textit{Phys. Lett.}  {\bf B 552}, 111 (2003);
G.~Shiu and I.~Wasserman, \textit{Phys. Lett}.  {\bf B 541}, 6 (2002);
D.~Choudhury et al., \textit{ Phys. Lett.}  {\bf B 544}, 231 (2002);
A.~V.~Frolov, et al., \textit{Phys. Lett.}  {\bf B 545}, 8 (2002);
G.~W.~Gibbons,\textit{ Phys. Lett.}  {\bf B 537}, 1 (2002).


\bibitem{ellis}
G.F.R.  Ellis  and M.S.Madsen, {\it Class.Quan.Grav.} {\bf 8}, 667 (1991); also see
F.E. Schunck, E. W. Mielke, Phys.Rev.\textbf{D50}, 4794 (1994).

\bibitem{jbp}
H.K. Jassal et al., \textit{MNRAS} \textbf{356}, L11-L16 (2005), [astro-ph/0404378]; [astro-ph/0601389].


	     
\bibitem{semicgrav}
V.G. Lapchinsky, V.A. Rubakov, \textit{ Acta Phys.Polon.} \textbf{B 10}, 1041 (1979);
J. B. Hartle, \textit{Phys. Rev.} \textbf{ D 37}, 2818 (1988);
               \textbf{D 38}, 2985 (1988);
T.Padmanabhan, \textit{Class. Quan. Grav. } \textbf{ 6}, 533 (1989);
T.P. Singh, T. Padmanabhan, \textit{Annals Phys.} \textbf{196}, 296(1989).
T. Padmanabhan, \textit{ Phys.Rev.} \textbf{D 39}, 2924 (1989);
T.Padmanabhan and T.P. Singh, Class. Quan. Grav.. \textbf{ 7}, 441 (1990);
J.J. Hallwell, \textit{Phys.Rev.} \textbf{D 39}, 2912 (1989).  
 	     
	     
	     
\bibitem{choices}
 T. Padmanabhan  and T. Roy Choudhury,  \textit{Mod. Phys. Lett.} \textbf{A 15}, 1813 (2000), [gr-qc/0006018].          
	     


\bibitem{zeropoint}
H. S. Snyder \textit{ Phys. Rev.}, \textbf{71}, 38 (1947);
B. S. DeWitt,\textit{ Phys. Rev. Lett.}, \textbf{13}, 114 (1964);
T. Yoneya  \textit{Prog. Theor. Phys.}, \textbf{56}, 1310 (1976);
T. Padmanabhan \textit{ Ann. Phys.} (N.Y.), \textbf{165}, 38 (1985); 
            \textit{Class. Quantum Grav.} \textbf{4}, L107 (1987);
A. Ashtekar et al., \textit{Phys. Rev. Lett.}, \textbf{69}, 237 (1992 );
T. Padmanabhan  \textit{Phys. Rev. Lett.} \textbf{78},  1854 (1997) [hep-th/9608182];
              \textit{Phys. Rev.}  \textbf{D 57}, 6206 (1998); 
K.Srinivasan et al., \textit{Phys. Rev. } \textbf{D 58} 044009 (1998) [gr-qc/9710104]; 
X.Calmet et al.,Phys.Rev.Lett.\textbf{93}:211101,(2004);hep-th/0505144;
M. Fontanini et al.  \textit{Phys.Lett.} \textbf{B 633}, 627 (2006)  hep-th/0509090.
For a review, see L.J. Garay, \textit{ Int. J. Mod. Phys.} \textbf{A10}, 145 (1995).

	     

\bibitem{ghds} 
G. W. Gibbons and S.W. Hawking, \textit{Phys. Rev.}  {\bf D 15},  2738 (1977);
T.Padmanabhan,\textit{Mod.Phys.Letts. }  \textbf{ A 19}, 2637 (2004) [gr-qc/0405072];



\bibitem{note1} Incidentally, $L_{DE}\approx 0.04$ mm is macroscopic; it is also pretty close to the length scale associated with a neutrino mass of $10^{-2}$ eV; another intriguing coincidence ?!



\bibitem{plumian}
T.Padmanabhan,  Lecture at the \textit{ Plumian 300 - The Quest for a Concordance Cosmology and Beyond } Institute of Astronomy, Cambridge, UK, July 2004;   [astro-ph/0510492].

\bibitem{bjorken} 
J.D. Bjorken, (2004)  astro-ph/0404233. 

\bibitem{cc1}

T. Padmanabhan, \textit{Gravity: A New Holographic Perspective } 
  (Lecture at the International Conference on Einstein's Legacy in the New Millennium, Puri, India, Dec, 2005)  Int.J.Mod.Phys., \textbf{D 15}, 1659-1675 (2006) [gr-qc/0606061].

   	    		 


\bibitem{cc2}  T. Padmanabhan Class.Quan.Grav., \textbf{22}, L107-L110, (2005) [hep-th/0406060]. For earlier attempts in similar spirit, see
T. Padmanabhan, \textit{Class.Quan.Grav.}  \textbf{19}, L167 (2002), [gr-qc/0204020]; 
D. Sorkin, \textit{ Int.J.Theor.Phys.} \textbf{36}, 2759 (1997); 
for related work, see
S.G.Djorgovski,V.G. Gurzadyan,astro-ph/0610204;
Volovik, G. E., gr-qc/0405012;
J. V. Lindesay et al., astro-ph/0412477;
Y. S. Myung, hep-th/0412224;
J.D.Barrow, gr-qc/0612128;
E.Elizalde et al., hep-th/0502082.


\bibitem{tpholo}
T. Padmanabhan, 
\textit{Brazilian Jour.Phys.} (Special Issue) \textbf{35}, 362 (2005) [gr-qc/0412068]; 
\textit{Class. Quan. Grav.}, \textbf{ 21}, L1 (2004) [gr-qc/0310027];
\textit{Gen.Rel.Grav.}, \textbf{35}, 2097 (2003); 
\textit{Gen. Rel. Grav.},\textbf{34}  2029 (2002) [gr-qc/0205090].
For related ideas, see e.g., 
E. Mottola, R. Vaulin,gr-qc/0604051; 
D. Cremades et al.,hep-th/0608174.

\bibitem{catchall} 
There is extensive literature on different paradigms for solving the cosmological constant problem,
 like e.g., those based on new symmetries:  
 R.Erdem, hep-th/0410063;
 G.'t Hooft, S.Nobbenhuis, gr-qc/0602076;
 D.E. Kaplan, R.Sundrum, hep-th/0505265;
 those based on QFT in CST:
 I. Antoniadis et al.,gr-qc/0612068;
 E. Mottola, \textit{Phys.Rev.} \textbf{D 31}, 754 (1985); 
N.C. Tsamis,R.P. Woodard, \textit{Phys.Lett.} \textbf{B301}, 351 (1993);
E. Elizalde  and S.D. Odintsov,  \textit{Phys.Lett.} {\bf B 321} 199 (1994); 
           {\bf B 333} 331 (1994);
J.V. Lindesay, H.P. Noyes, astro-ph/0508450;
S. Wang, gr-qc/0606109;
Shi Qi, hep-th/0505109.
Non-ideal fluids mimicking cosmological constant, like e.g.,
 T.Padmanabhan  and S.M. Chitre, \textit{Phys. Letts.} \textbf{A 120}, 433 (1987);
 Quantum cosmological considerations: 
E. Baum, \textit{Phys. Letts.}  {\bf  B 133},  185 (1983);
T. Padmanabhan,  \textit{Phys. Letts.},\textbf{ A104}, 196  (1984); 
S.W. Hawking, \textit{Phys. Letts.}  {\bf  B 134},  403 (1984); 
 S. Coleman, \textit{Nucl. Phys.} {\bf B 310}  643 (1988);
T. Mongan, \textit{ Gen. Rel. Grav.}, {\bf 33} 1415 (2001) [gr-qc/0103021]; 
        \textit{Gen.Rel.Grav.} \textbf{35}  685 (2003).
Holographic dark energy:
Qiang Wu et al.,arXiv:0705.1006;
Y. S. Myung, gr-qc/0702032;
F. Simpson, astro-ph/0609755;
Y. Gong, Yuan-Zhong Zhang, hep-th/0505175; 
         astro-ph/0502262; 
	 hep-th/0412218;
 X. Zhang, Feng-Quan Wu, astro-ph/0506310;
 M.G.Hu,  Xin-He Meng, astro-ph/0511615.
Those based on renormalization group, running coupling constants and more general time dependent
decay schemes:
I.L. Shapiro,J.Sola,Phys.Lett.\textbf{B574}:149-155 (2003) [astro-ph/0303306];
I.L.  Shapiro and J. Sola, \textit{Phys.Lett.} {\bf B475}, 236 (2000); 
I.L. Shapiro et al., hep-ph/0410095;
C. Espana-Bonet, et.al., \textit{Phys.Lett.} \textbf{B 574} 149 (2003); 
        \textit{JCAP} \textbf{0402}, 006 (2004);
F.Bauer, gr-qc/0501078; 
         gr-qc/0512007;
I. L. Shapiro, J. Sola, \textit{JHEP}, \textbf{ 0202 } 006 (2002);
J.Sola and H.Stefancic, astro-ph/0505133;
J. S. Alcaniz, J. A. S. Lima, \textit{Phys.Rev.} \textbf{ D 72} 063516 (2005), [astro-ph/0507372];
J. S. Alcaniz , J. M. F. Maia, \textit{Phys.Rev.} \textbf{D 67}  043502 (2003);
F. Bauer, \textit{Class.Quant.Grav.}, \textbf{22}, 3533 (2005) and many more. 

\bibitem{gofemw}  
T.Padmanabhan,  Gen. Rel. Grav. , (1987), \textbf{ 19} , 927.
 
\bibitem{flucde} 
See, for example,
T.Buchert, gr-qc/0612166; 
          Gen.Rel.Grav.\textbf{32}, 105,(2000);
	  Gen.Rel.Grav.\textbf{33}, 1381 (2001);
G.F.R. Ellis, T. Buchert, Phys.Lett.\textbf{A347}, 38 (2005);
R.M. Zalaletdinov,Gen.Rel.Grav.\textbf{24}, 1015 (1992);
A. Paranjape, T. P. Singh, gr-qc/0703106; 
                 Class.Quant.Grav. \textbf{23} (2006) 6955;
A. Ishibashi, R.M. Wald, Class.Quant.Grav. \textbf{23} (2006) 235-250;
S.Rasanen, JCAP 0402 (2004) 003;
B. Losic, W.G. Unruh, Phys.Rev.\textbf{D72}:123510, (2005);
C. M. Hirata, U. Seljak, Phys.Rev.\textbf{D72}:083501, (2005);
E.W. Kolb et al.,astro-ph/0506534;  hep-th/0503117.
 
 

\bibitem{Hbubble}
 S. Jha et al.,  Astrophys.J. \textbf{659} (2007) 122;
 A. Conley et al., arXiv:0705.0367v1.

 

\bibitem{lenz}    
A.D. Dolgov,  in {\it The very early universe: Proceeding of the 1982 Nuffield  Workshop at Cambridge}, ed. G.W. Gibbons, S.W. Hawking and S.T.C. Sikkos (Cambridge University Press), (1982), p. 449; 
S.M. Barr, \textit{Phys. Rev.} {\bf D 36}, 1691 (1987); 
Ford, L.H.,\textit{ Phys. Rev. }{\bf D 35},  2339 (1987); 
A. Hebecker  and C. Wetterich,  \textit{Phy. Rev. Lett.}, {\bf 85} 3339 (2000); 
A. Hebecker   hep-ph/0105315; 
T.P.  Singh,  T. Padmanabhan, \textit{Int. Jour. Mod. Phys.} {\bf A 3}, (1988), 1593;
M. Sami,   T. Padmanabhan,  (2003) \textit{Phys. Rev.}   \textbf{D 67}, 083509 [hep-th/0212317].




\bibitem{unimod} 
A. Einstein, \textit{Siz. Preuss. Acad. Scis.} (1919), translated in The \textit{Principle of Relativity}, by edited by A. Einstein et al. (Dover, New York, 1952); 
J. J. van der Bij et al., \textit{Physica} \textbf{A 116}, 307 (1982);
F. Wilczek, \textit{Phys. Rep.} \textbf{104}, 111 (1984); 
A. Zee, in \textit{High Energy Physics}, proceedings of the 20th Annual Orbis Scientiae, Coral Gables, (1983), edited by B. Kursunoglu, S. C. Mintz, and A. Perlmutter (Plenum, New York, 1985); 
W. Buchmuller and N. Dragon,  \textit{Phys.Lett.} {\bf B 207}, 292, (1988); 
W.G. Unruh, \textit{Phys.Rev.}  {\bf D 40} 1048 (1989).   

\bibitem{probe}
S A Fulling, \textit{Phys. Rev.} \textbf{D7}, 2850 (1973);
W.G. Unruh, \textit{ Phys. Rev.} \textbf{D14}, 870 (1976);
B.S. DeWitt,  in  {\it General Relativity: An Einstein Centenary Survey}, pp 680-745
 Cambridge University Press, (1979),ed., S.W. Hawking and W. Israel;
T. Padmanabhan,Class. Quan. Grav. , (1985), \textbf{2} , 117; 
K. Srinivasan et al.,\textit{ Phys. Rev.} \textbf{D 60}, 24007 (1999) [gr-qc/9812028];
L Sriramkumar et al., \textit{Int. Jour. Mod. Phys.},  \textbf{ D
  11},1 (2002) [gr-qc/9903054].


\bibitem{gravitonmyth} T. Padmanabhan, \textit{From gravitons to gravity: myths and  reality}, (2004), [gr-qc/0409089]

\bibitem{gr06} T. Padmanabhan, \textit{Gravity's Immunity from Vacuum: The  Holographic Structure  of Semiclassical Action},
Third prize essay; Gravity Essay Contest 2006 
\textit{Gen.Rel.Grav}., \textbf{ 38}, 1547-1552 (2006);
\textit{ Int.J.Mod.Phys.}, \textbf{D 15},  2029 (2006) [gr-qc/0609012]

\bibitem{gaurang}
G. Mahajan et al.,astro-ph/0604265.

\bibitem{elastic}
A. D. Sakharov,  \textit{Sov. Phys. Dokl.}, {\bf 12}, 1040 (1968);
T.~Jacobson,  \textit{Phys. Rev. Lett.} \textbf{75},  1260 (1995);
T.~Padmanabhan, \textit{Mod. Phys. Lett.} \textbf{A 17}, 1147   (2002)[hep-th/0205278]; 
                \textit{Int.Jour.Mod.Phys.} \textbf{D 13}, 2293-2298 (2004), [gr-qc/0408051];
                \textbf{18}, 2903 (2003) [hep-th/0302068];
              \textit{Class.Quan.Grav.}, \textbf{21}, 4485 (2004) [gr-qc/0308070];
G.E. Volovik, \textit{Phys.Rept.}, \textbf{351}, 195 (2001); 
G.E. Volovik, gr-qc/0604062; 
G.~E. Volovik, \textit{The universe in a helium droplet}, (Oxford University Press, 2003); 
Chao-Guang Huang, Jia-Rui Sun,gr-qc/0701078;
J.Makela, gr-qc/0701128.  



\bibitem{magglass}
 See e.g.,T.Padmanabhan, Phys. Rev. Letts.,\textbf{81},4297 (1998)[hep-th/9801015]; 
                   Phys. Rev. D., 59, 124012 (1999) [hep-th/9801138]
 and references therein. 


  
  
\bibitem{paddy1}
For a recent review, see:
T Padmanabhan, {\it Phys. Rept.}, {\bf 406}, 49, (2005),[gr-qc/0311036]; 
               {\it Mod. Phys. Lett.} {\bf A17}, 923, (2002) [gr-qc/0202078]. 




\bibitem{paddy2}
T Padmanabhan,{\it Class. Quan. Grav.}, {\bf 19}, 5387, (2002) [gr-qc/0204019]. 


\bibitem{rongencai}
J. Zhou et al.,arXiv:0705.1264;
R-G Cai, L-M Cao, [gr-qc/0611071]; 
M Akbar, R-G Cai, [hep-th/0609128];
M. Akbar, [hep-th/0702029]

\bibitem{dawood-sudipta-tp} 
D. Kothawala, S. Sarkar, T Padmanabhan, \textit{Einstein's
  equations as a thermodynamic identity: The cases of stationary
  axis-symmetric horizons and evolving spherically symmetric
  horizons}, gr-qc/0701002. 
  
  
\bibitem{paris}  T. Padmanabhan, Plenary talk at the  Albert Einstein Century International Conference, Paris, 18-22 July 2005, AIP Conference Proceedings  \textbf{861}, 858-866, [astro-ph/0603114];
\textit{Int.Jour.Mod.Phys} \textbf{ D14}, 2263-2270 (2005) [gr-qc/0510015].


\bibitem{aseem-sudipta}
A Paranjape, S Sarkar and T Padmanabhan, {\it Phys. Rev.} {\bf D74},
                          104015, (2006) [hep-th/0607240].  

\bibitem{cai2} 
M. Akbar, Rong-Gen Cai, gr-qc/0612089; 
Xian-Hui Ge,hep-th/0703253;
A. Sheykhi et al.,hep-th/0701198;
G. Allemandi et al.,gr-qc/0308019
  
\bibitem{ayan}
A. Mukhopadhyay, T. Padmanabhan, Phys.Rev., \textbf{D 74}, 124023 (2006) [hep-th/0608120]  


\bibitem{aseementropy} This is a  generalisation of the ideas presented in an earlier work, which only considered null normals:
 T. Padmanabhan, A. Paranjape, \textit{Phys.Rev.} \textbf{D75} 064004, (2007) [gr-qc/0701003].  


 \bibitem{lovelock} 
C Lanczos, {\it Z. Phys.} {\bf 73}, 147, (1932); {\it Annals Math.} {\bf 39}, 842, (1938); 
D Lovelock, {\it J. Math. Phys.}, {\bf 12}, 498 (1971). 

\bibitem{gh}
J.W. York, Phys.Rev.Letts., \textbf{28} (1972) 1082;
G.~W. Gibbons, S.~W. Hawking,  Phys. Rev. \textbf{D15} (1977) 2752--2756


\bibitem{noether}
 R. M. Wald, Phys. Rev. D \textbf{48}, 3427 (1993);
 V. Iyer, R. M. Wald, Phys. Rev. D \textbf{52}, 4430-4439 (1995).

\bibitem{note3}
On shell, the last two terms in the action in \eq{on-shell-1} is the same as the conventional one for gravity coupled matter, if $\epsilon=1$ but the surface term in, for example, \eq{ygh} has the wrong sign. 
 
 
\bibitem{tptptmunu}
T. Padmanabhan and T.P. Singh,  \textit{Class. Quan. Grav}., (1987), \textbf{4} , 1397. 

\bibitem{area}
L.~Bombelli et al., \textit{Phys. Rev.} {\bf D34}, 373 (1986);
M.~Srednicki, \textit{Phys. Rev. Lett.} {\bf 71}, 666 (1993);
R.~Brustein et al., \textit{Phys. Rev.} {\bf D65}, 105013 (2002);
A. Yarom, R. Brustein, hep-th/0401081. 
This result can also be obtained from those in ref. \cite{tptptmunu}.


	       
\end{thebibliography}
 \end{document}